\documentclass[11pt]{article}%\pdfoutput=1
\usepackage{hyperref} % allows to link references, citations, etc
\usepackage{cite}
\usepackage{scalerel}
\usepackage{amsmath,amsfonts,amssymb}
\usepackage{latexsym,epsfig}

\usepackage{extarrows}

\usepackage{tikz}
\usetikzlibrary{positioning,arrows,calc}
\usetikzlibrary{arrows.meta}
\usetikzlibrary{decorations.pathmorphing}
\usetikzlibrary{decorations.markings}
\usetikzlibrary{shapes.geometric}
\usetikzlibrary{patterns}
\usepackage{xparse}
\usetikzlibrary{shapes.misc} % new shapes for tikz, like cross.
\newif\ifstartcompletesineup
\newif\ifendcompletesineup
\pgfkeys{
    /pgf/decoration/.cd,
    start up/.is if=startcompletesineup,
    start up=true,
    start up/.default=true,
    start down/.style={/pgf/decoration/start up=false},
    end up/.is if=endcompletesineup,
    end up=true,
    end up/.default=true,
    end down/.style={/pgf/decoration/end up=false}
}
\pgfdeclaredecoration{complete sines}{initial}
{
    \state{initial}[
        width=+0pt,
        next state=upsine,
        persistent precomputation={
            \ifstartcompletesineup
                \pgfkeys{/pgf/decoration automaton/next state=upsine}
                \ifendcompletesineup
                    \pgfmathsetmacro\matchinglength{
                        0.5*\pgfdecoratedinputsegmentlength / (ceil(0.5* \pgfdecoratedinputsegmentlength / \pgfdecorationsegmentlength) )
                    }
                \else
                    \pgfmathsetmacro\matchinglength{
                        0.5 * \pgfdecoratedinputsegmentlength / (ceil(0.5 * \pgfdecoratedinputsegmentlength / \pgfdecorationsegmentlength ) - 0.499)
                    }
                \fi
            \else
                \pgfkeys{/pgf/decoration automaton/next state=downsine}
                \ifendcompletesineup
                    \pgfmathsetmacro\matchinglength{
                        0.5* \pgfdecoratedinputsegmentlength / (ceil(0.5 * \pgfdecoratedinputsegmentlength / \pgfdecorationsegmentlength ) - 0.4999)
                    }
                \else
                    \pgfmathsetmacro\matchinglength{
                        0.5 * \pgfdecoratedinputsegmentlength / (ceil(0.5 * \pgfdecoratedinputsegmentlength / \pgfdecorationsegmentlength ) )
                    }
                \fi
            \fi
            \setlength{\pgfdecorationsegmentlength}{\matchinglength pt}
        }] {}
    \state{downsine}[width=\pgfdecorationsegmentlength,next state=upsine]{
        \pgfpathsine{\pgfpoint{0.5\pgfdecorationsegmentlength}{0.5\pgfdecorationsegmentamplitude}}
        \pgfpathcosine{\pgfpoint{0.5\pgfdecorationsegmentlength}{-0.5\pgfdecorationsegmentamplitude}}
    }
    \state{upsine}[width=\pgfdecorationsegmentlength,next state=downsine]{
        \pgfpathsine{\pgfpoint{0.5\pgfdecorationsegmentlength}{-0.5\pgfdecorationsegmentamplitude}}
        \pgfpathcosine{\pgfpoint{0.5\pgfdecorationsegmentlength}{0.5\pgfdecorationsegmentamplitude}}
}
    \state{final}{}
}
\definecolor{inta}{RGB}{207,54,108}
\definecolor{intb}{RGB}{53,152,219}
\definecolor{ext}{RGB}{176,156,255}
\tikzset{
L/.style={line width=1pt,draw=red},
D/.style={line width=1pt,densely dotted,draw=blue},
E/.style={line width=1pt,decorate, draw=blue,
	decoration={complete sines,aspect=0,amplitude=1mm,segment length=1mm,start up,end up}},
X/.style={cross out, draw=black, line width = 1pt, minimum size=4*(#1-\pgflinewidth), inner sep=1.2pt, outer sep=1.2pt}, 
cross/.default={3pt}
}

\newcommand\scale{.7} % loop scale factor
\newcommand\scalev{1} % vertex scale factor
\newcommand\disex{.5} % distance of external lines
\newcommand\rad{.5} % Radius of the loops
\newcommand\cx{3} %center of x
\newcommand\cy{.5} %center of y
\newcommand\circsize{.05} %size of vertex circles
\newcommand\Xsize{.12} %size of counterterm cross

\newcommand{\ZeroVa}[3]{\scalebox{\scalev}{\begin{tikzpicture}[baseline={([yshift=-.5ex]current bounding box.center)}]
		
		\draw[#1] (\cx,\cy) -- (\cx+\disex,\cy+.4);
		\draw[#2] (\cx,\cy) -- (\cx+\disex,\cy);
		\draw[#3] (\cx,\cy) -- (\cx+\disex,\cy-.4);
		
		\draw[fill=white] (\cx,\cy) circle (\circsize);
		
		\end{tikzpicture}}}

\newcommand{\ZeroVb}[4]{\scalebox{\scalev}{\begin{tikzpicture}[baseline={([yshift=-.5ex]current bounding box.center)}]
		
		\draw[#1] (\cx,\cy) -- (\cx+\disex,\cy+.4);
		\draw[#2] (\cx,\cy) -- (\cx+\disex,\cy+.2);
		\draw[#3] (\cx,\cy) -- (\cx+\disex,\cy-.2);
		\draw[#4] (\cx,\cy) -- (\cx+\disex,\cy-.4);
		
		\draw[fill=white] (\cx,\cy) circle (\circsize);
		
		\end{tikzpicture}}}
	
\newcommand{\OneVa}[2]{\scalebox{\scalev}{\begin{tikzpicture}[baseline={([yshift=-.5ex]current bounding box.center)}]
		
		\draw[E]  (\cx-\disex,\cy) -- (\cx,\cy);
		\draw[#1] (\cx,\cy) -- (\cx+\disex,\cy+.4);
		\draw[#2] (\cx,\cy) -- (\cx+\disex,\cy-.4);
		
		\draw[fill=white] (\cx,\cy) circle (\circsize);
		
		\end{tikzpicture}}}
	
\newcommand{\OneVb}[3]{\scalebox{\scalev}{\begin{tikzpicture}[baseline={([yshift=-.5ex]current bounding box.center)}]
		
		\draw[E]  (\cx-\disex,\cy) -- (\cx,\cy);
		\draw[#1] (\cx,\cy) -- (\cx+\disex,\cy+.4);
		\draw[#2] (\cx,\cy) -- (\cx+\disex,\cy);
		\draw[#3] (\cx,\cy) -- (\cx+\disex,\cy-.4);
		
		\draw[fill=white] (\cx,\cy) circle (\circsize);
		
		\end{tikzpicture}}}
	
\newcommand{\OneVc}[4]{\scalebox{\scalev}{\begin{tikzpicture}[baseline={([yshift=-.5ex]current bounding box.center)}]
		
		\draw[E]  (\cx-\disex,\cy) -- (\cx,\cy);
		\draw[#1] (\cx,\cy) -- (\cx+\disex,\cy+.4);
		\draw[#2] (\cx,\cy) -- (\cx+\disex,\cy+.2);
		\draw[#3] (\cx,\cy) -- (\cx+\disex,\cy-.2);
		\draw[#4] (\cx,\cy) -- (\cx+\disex,\cy-.4);
		
		\draw[fill=white] (\cx,\cy) circle (\circsize);
		
		\end{tikzpicture}}}
	
\newcommand{\TwoVa}[2]{\scalebox{\scalev}{\begin{tikzpicture}[baseline={([yshift=-.5ex]current bounding box.center)}]
		
		\draw[E]  (\cx-\disex,\cy+.4) -- (\cx,\cy);
		\draw[E]  (\cx-\disex,\cy-.4) -- (\cx,\cy);
		\draw[#1] (\cx,\cy) -- (\cx+\disex,\cy+.4);
		\draw[#2] (\cx,\cy) -- (\cx+\disex,\cy-.4);
		
		\draw[fill=white] (\cx,\cy) circle (\circsize);
		
		\end{tikzpicture}}}

\newcommand{\TwoVb}[3]{\scalebox{\scalev}{\begin{tikzpicture}[baseline={([yshift=-.5ex]current bounding box.center)}]
		
		\draw[E]  (\cx-\disex,\cy+.4) -- (\cx,\cy);
		\draw[E]  (\cx-\disex,\cy-.4) -- (\cx,\cy);
		\draw[#1] (\cx,\cy) -- (\cx+\disex,\cy+.4);
		\draw[#2] (\cx,\cy) -- (\cx+\disex,\cy);
		\draw[#3] (\cx,\cy) -- (\cx+\disex,\cy-.4);
		
		\draw[fill=white] (\cx,\cy) circle (\circsize);
		
		\end{tikzpicture}}}

\newcommand{\TwoVc}[4]{\scalebox{\scalev}{\begin{tikzpicture}[baseline={([yshift=-.5ex]current bounding box.center)}]
		
		\draw[E]  (\cx-\disex,\cy+.4) -- (\cx,\cy);
		\draw[E]  (\cx-\disex,\cy-.4) -- (\cx,\cy);
		\draw[#1] (\cx,\cy) -- (\cx+\disex,\cy+.4);
		\draw[#2] (\cx,\cy) -- (\cx+\disex,\cy+.2);
		\draw[#3] (\cx,\cy) -- (\cx+\disex,\cy-.2);
		\draw[#4] (\cx,\cy) -- (\cx+\disex,\cy-.4);
		
		\draw[fill=white] (\cx,\cy) circle (\circsize);
		
		\end{tikzpicture}}}

\newcommand{\OneLoopA}[6]{\scalebox{\scale}{\begin{tikzpicture}[baseline={([yshift=-.5ex]current bounding box.center)}]
		
		\draw[#1] (\cx-\disex,\cy-.2) -- (\cx,\cy);
		\draw[#2] (\cx+2*\rad,\cy) -- (\cx+2*\rad+\disex,\cy-.2);
		\draw[#3] (\cx-\disex,\cy+.2) -- (\cx,\cy);
		\draw[#4] (\cx+2*\rad,\cy) -- (\cx+2*\rad+\disex,\cy+.2);
		
		\draw[#5] (\cx,\cy) arc (180:0:\rad);
		\draw[#6] (\cx,\cy) arc (-180:0:\rad);
		
		\draw[fill=white] (\cx,\cy) circle (\circsize);
		\draw[fill=white] (\cx+2*\rad,\cy) circle (\circsize);
		
		\end{tikzpicture}}}

\newcommand{\OneLoopI}[3]{\scalebox{\scale}{\begin{tikzpicture}[baseline={([yshift=-.5ex]current bounding box.center)}]
		
		\draw[E] (\cx+\rad,\cy-\rad) -- (\cx+\rad-.4,\cy-\rad-\disex);
		\draw[E] (\cx+\rad,\cy-\rad) -- (\cx+\rad+.4,\cy-\rad-\disex);
		\draw[#1] (\cx+\rad,\cy-\rad) -- (\cx+\rad-.2,\cy-\rad-\disex);
		\draw[#2] (\cx+\rad,\cy-\rad) -- (\cx+\rad+.2,\cy-\rad-\disex);
		
		\draw[#3] (\cx,\cy) arc (180:0:\rad);
		\draw[#3] (\cx,\cy) arc (-180:0:\rad);
		
		\draw[fill=white] (\cx+\rad,\cy-\rad) circle (\circsize);
\end{tikzpicture}}}

\newcommand{\CounterLoopA}[1]{\scalebox{\scale}{\begin{tikzpicture}[baseline={([yshift=-.5ex]current bounding box.center)}]
		
		\draw[E] (\cx+\rad,\cy-\rad) -- (\cx+\rad-.4,\cy-\rad-\disex);
		\draw[E] (\cx+\rad,\cy-\rad) -- (\cx+\rad+.4,\cy-\rad-\disex);
		
		\draw[#1] (\cx,\cy) arc (180:0:\rad);
		\draw[#1] (\cx,\cy) arc (-180:0:\rad);
		
		\draw[fill=white] (\cx+\rad,\cy-\rad) circle (\Xsize);
		\draw (\cx+\rad,\cy-\rad) node[X]{};
		
		\end{tikzpicture}}}

\newcommand{\TwoLoopA}[4]{\scalebox{\scale}{\begin{tikzpicture}[baseline={([yshift=-.5ex]current bounding box.center)}]
		
		\draw[E] (\cx+2*\rad,\cy) -- (\cx+2*\rad,\cy+\disex+.2);
		\draw[E] (\cx+2*\rad,\cy) -- (\cx+2*\rad,\cy-\disex-.2);
		
		\draw[#1] (\cx,\cy) arc (180:0:\rad);
		\draw[#2] (\cx,\cy) arc (-180:0:\rad);
		\draw[#3] (\cx+2*\rad,\cy) arc (180:0:\rad);
		\draw[#4] (\cx+2*\rad,\cy) arc (-180:0:\rad);
		
		\draw[fill=white] (\cx+2*\rad,\cy) circle (\circsize);
		
		\end{tikzpicture}}}

\newcommand{\TwoLoopB}[4]{\scalebox{\scale}{\begin{tikzpicture}[baseline={([yshift=-.5ex]current bounding box.center)}]
		
		\draw[E] (\cx+2*\rad,\cy) -- (\cx+2*\rad,\cy+\disex+.2);
		\draw[E] (\cx+4*\rad,\cy) -- (\cx+4*\rad+\disex,\cy);
		
		\draw[#1] (\cx,\cy) arc (180:0:\rad);
		\draw[#2] (\cx,\cy) arc (-180:0:\rad);
		\draw[#3] (\cx+2*\rad,\cy) arc (180:0:\rad);
		\draw[#4] (\cx+2*\rad,\cy) arc (-180:0:\rad);
		
		\draw[fill=white] (\cx+2*\rad,\cy) circle (\circsize);
		\draw[fill=white] (\cx+4*\rad,\cy) circle (\circsize);
				
		\end{tikzpicture}}}
	
\newcommand{\TwoLoopC}[4]{\scalebox{\scale}{\begin{tikzpicture}[baseline={([yshift=-.5ex]current bounding box.center)}]
		
		\draw[E] (\cx+4*\rad,\cy) -- (\cx+4*\rad+\disex,\cy+.2);
		\draw[E] (\cx+4*\rad,\cy) -- (\cx+4*\rad+\disex,\cy-.2);
		
		\draw[#1] (\cx,\cy) arc (180:0:\rad);
		\draw[#2] (\cx,\cy) arc (-180:0:\rad);
		\draw[#3] (\cx+2*\rad,\cy) arc (180:0:\rad);
		\draw[#4] (\cx+2*\rad,\cy) arc (-180:0:\rad);
		
		\draw[fill=white] (\cx+2*\rad,\cy) circle (\circsize);
		\draw[fill=white] (\cx+4*\rad,\cy) circle (\circsize);
		
		\end{tikzpicture}}}
	
\newcommand{\TwoLoopD}[4]{\scalebox{\scale}{\begin{tikzpicture}[baseline={([yshift=-.5ex]current bounding box.center)}]
		
		\draw[E] (\cx+3*\rad,\cy+\rad) -- (\cx+3*\rad,\cy+\rad+\disex);
		\draw[E] (\cx+3*\rad,\cy-\rad) -- (\cx+3*\rad,\cy-\rad-\disex);
		
		\draw[#1] (\cx,\cy) arc (180:0:\rad);
		\draw[#2] (\cx,\cy) arc (-180:0:\rad);
		\draw[#3] (\cx+3*\rad,\cy+\rad) arc (90:270:\rad);
		\draw[#4] (\cx+3*\rad,\cy-\rad) arc (-90:90:\rad);

		\draw[fill=white] (\cx+2*\rad,\cy) circle (\circsize);
		\draw[fill=white] (\cx+3*\rad,\cy+\rad) circle (\circsize);
		\draw[fill=white] (\cx+3*\rad,\cy-\rad) circle (\circsize);
		
		\end{tikzpicture}}}
	
\newcommand{\TwoLoopE}[4]{\scalebox{\scale}{\begin{tikzpicture}[baseline={([yshift=-.5ex]current bounding box.center)}]
		
		\draw[E] (\cx-\disex,\cy) -- (\cx,\cy);
		\draw[E] (\cx+4*\rad,\cy) -- (\cx+4*\rad+\disex,\cy);
		
		\draw[#1] (\cx,\cy) arc (180:0:\rad);
		\draw[#2] (\cx,\cy) arc (-180:0:\rad);
		\draw[#3] (\cx+2*\rad,\cy) arc (180:0:\rad);
		\draw[#4] (\cx+2*\rad,\cy) arc (-180:0:\rad);
		
		\draw[fill=white] (\cx,\cy) circle (\circsize);
		\draw[fill=white] (\cx+2*\rad,\cy) circle (\circsize);
		\draw[fill=white] (\cx+4*\rad,\cy) circle (\circsize);
		
		\end{tikzpicture}}}
	
\newcommand{\TwoLoopF}[3]{\scalebox{\scale}{\begin{tikzpicture}[baseline={([yshift=-.5ex]current bounding box.center)}]
		
		\draw[E] (\cx-\disex,\cy) -- (\cx,\cy);
		\draw[E] (\cx+2*\rad,\cy) -- (\cx+2*\rad+\disex,\cy);
		
		\draw[#1] (\cx,\cy) arc (180:0:\rad);
		\draw[#2] (\cx,\cy) arc (-180:0:\rad);
		\draw[#3] (\cx,\cy) -- (\cx+2*\rad,\cy);
		
		\draw[fill=white] (\cx,\cy) circle (\circsize);
		\draw[fill=white] (\cx+2*\rad,\cy) circle (\circsize);
		
		\end{tikzpicture}}}
		
\newcommand{\TwoLoopG}[3]{\scalebox{\scale}{\begin{tikzpicture}[baseline={([yshift=-.5ex]current bounding box.center)}]
		
		\draw[E] (\cx+\rad,\cy+\rad) -- (\cx+\rad,\cy+\rad+\disex);
		\draw[E] (\cx+2*\rad,\cy) -- (\cx+2*\rad+\disex,\cy);
		
		\draw[#1] (\cx+\rad,\cy+\rad) arc (90:360:\rad);
		\draw[#2] (\cx+2*\rad,\cy) arc (0:90:\rad);
		\draw[#3] (\cx,\cy) -- (\cx+2*\rad,\cy);
		
		\draw[fill=white] (\cx,\cy) circle (\circsize);
		\draw[fill=white] (\cx+2*\rad,\cy) circle (\circsize);
		\draw[fill=white] (\cx+\rad,\cy+\rad) circle (\circsize);
		
		\end{tikzpicture}}}
	
\newcommand{\TwoLoopH}[3]{\scalebox{\scale}{\begin{tikzpicture}[baseline={([yshift=-.5ex]current bounding box.center)}]
		
		\draw[E] (\cx+\rad,\cy-\rad) -- (\cx+\rad+.2,\cy-\rad-\disex);
		\draw[E] (\cx+\rad,\cy-\rad) -- (\cx+\rad-.2,\cy-\rad-\disex);
		
		\draw[#1] (\cx,\cy) arc (180:0:\rad);
		\draw[#2] (\cx,\cy) arc (-180:0:\rad);
		\draw[#3] (\cx,\cy) -- (\cx+2*\rad,\cy);
		
		\draw[fill=white] (\cx,\cy) circle (\circsize);
		\draw[fill=white] (\cx+2*\rad,\cy) circle (\circsize);
		\draw[fill=white] (\cx+\rad,\cy-\rad) circle (\circsize);
		
		\end{tikzpicture}}}

\newcommand{\TwoLoopI}[5]{\scalebox{\scale}{\begin{tikzpicture}[baseline={([yshift=-.5ex]current bounding box.center)}]
		
		\draw[E] (\cx+\rad*.5,\cy-\rad*.87) -- (\cx+\rad*.5-.2,\cy-\rad*.87-\disex);
		\draw[E] (\cx+1.5*\rad,\cy-\rad*.87) -- (\cx+1.5*\rad+.2,\cy-\rad*.87-\disex);
		
		\draw[#1] (\cx,\cy) arc (180:0:\rad);
		\draw[#2] (\cx,\cy) arc (-180:-120:\rad);
		\draw[#3] (\cx+\rad*.5,\cy-\rad*.87) arc (-120:-60:\rad);
		\draw[#4] (\cx+\rad*1.5,\cy-\rad*.87) arc (-60:0:\rad);
		\draw[#5] (\cx,\cy) -- (\cx+2*\rad,\cy);
		
		\draw[fill=white] (\cx,\cy) circle (\circsize);
		\draw[fill=white] (\cx+2*\rad,\cy) circle (\circsize);
		\draw[fill=white] (\cx+\rad*.5,\cy-\rad*.87) circle (\circsize);
		\draw[fill=white] (\cx+1.5*\rad,\cy-\rad*.87) circle (\circsize);
				
		\end{tikzpicture}}}
	
\newcommand{\TwoLoopJ}[5]{\scalebox{\scale}{\begin{tikzpicture}[baseline={([yshift=-.5ex]current bounding box.center)}]
		
		\draw[E] (\cx+\rad,\cy+\rad) -- (\cx+\rad,\cy+\rad+\disex);
		\draw[E] (\cx+\rad,\cy-\rad) -- (\cx+\rad,\cy-\rad -\disex);
		
		\draw[#1] (\cx,\cy) arc (180:90:\rad);
		\draw[#2] (\cx+\rad,\cy+\rad) arc (90:0:\rad);
		\draw[#3] (\cx+2*\rad,\cy) arc (0:-90:\rad);
		\draw[#4] (\cx+\rad,\cy-\rad) arc (-90:-180:\rad);
		\draw[#5] (\cx,\cy) -- (\cx+2*\rad,\cy);
		
		\draw[fill=white] (\cx,\cy) circle (\circsize);
		\draw[fill=white] (\cx+2*\rad,\cy) circle (\circsize);
		\draw[fill=white] (\cx+\rad,\cy+\rad) circle (\circsize);
		\draw[fill=white] (\cx+\rad,\cy-\rad) circle (\circsize);
		
		\end{tikzpicture}}}

\usepackage{comment} % To comment several lines at once.

\def\hybrid{
        \topmargin -20pt
        \oddsidemargin 0pt
        \headheight 0pt \headsep 0pt
        \textwidth 6.25in % A4 paper
        \textheight 9.5in % A4 paper
        \marginparwidth .875in
        \parskip 5pt plus 1pt \jot = 1.5ex}

\hybrid

 \csname
@addtoreset\endcsname{equation}{section}

\def\cL{{\cal L}}
\def\cD{\mathcal{D}}

\def\cO{{\cal O}}
\def\cA{{\cal A}}

\def\cE{{\cal E}}
\def\cM{{\cal M}}

\def\cR{{\cal R}}
\def\cS{{\cal S}}

\def\cK{{\cal K}}

\def\cH{{\cal H}}

\def\cT{{\cal T}}
\def\cW{{\cal W}}
\def\del{\partial}
\def\l{\langle}
\def\r{\rangle}
\def\tr{{\rm tr}}
\def\Tr{{\rm Tr}}
\def\dk{\frac{d^nk}{(2\pi)^n}}

\allowdisplaybreaks
\def\K{\mathcal{K}}
\def\R{\mathcal{R}}

\begin{document}

\begin{titlepage}
\rightline{}
%\rightline\today
\rightline{November   2021}
\rightline{HU-EP-21/44-RTG}  
\begin{center}
\vskip 1.5cm
 {\Large \bf{Duality invariant string beta functions at two loops}}
\vskip 1.7cm

{\large\bf {Roberto Bonezzi, Tomas Codina and Olaf Hohm}}
\vskip 1.6cm

{\it  Institute for Physics, Humboldt University Berlin,\\
 Zum Gro\ss en Windkanal 6, D-12489 Berlin, Germany}\\[1.5ex] 
 ohohm@physik.hu-berlin.de, 
roberto.bonezzi@physik.hu-berlin.de, 
tomas.codina@physik.hu-berlin.de
\vskip .1cm

\vskip .2cm

\end{center}

\bigskip\bigskip
\begin{center} 
\textbf{Abstract}

\end{center} 
\begin{quote}

We compute, for cosmological backgrounds, the $O(d,d;\mathbb{R})$ invariant beta functions for 
the sigma model of  the bosonic string at two loops. 
This yields an independent  first-principle derivation of the order $\alpha'$ 
corrections to the cosmological target-space equations. 
To this end we revisit the quantum consistency of Tseytlin's duality invariant formulation of the worldsheet theory. 
While we confirm the absence of gravitational (and hence Lorentz) anomalies, our results 
show that the minimal subtraction scheme is not applicable, implying significant technical complications at 
higher loops. To circumvent these we then change gears and use the Polyakov action for 
cosmological backgrounds, applying a suitable  perturbation scheme that, although not $O(d,d;\mathbb{R})$ invariant, 
allows one to efficiently determine the $O(d,d;\mathbb{R})$ invariant beta functions.

\end{quote} 
\vfill
\setcounter{footnote}{0}
\end{titlepage}

\tableofcontents
%\newpage

\section{Introduction}

In this paper, which is a sequel to \cite{Bonezzi:2021mub}, our goal is twofold: First, 
we continue the program of computing 
duality invariant beta functions for the string sigma model. 
Second, we revisit (and reaffirm) the quantum consistency 
of the duality invariant sigma model due to Tseytlin \cite{Tseytlin:1990va} and its generalizations  \cite{Schwarz:1993mg,Blair:2016xnn,Bonezzi:2020ryb}. 
Here we refer by duality to the phenomenon that string theory with 
$d$-dimensional translation invariance admits an $O(d,d;\mathbb{R})$ symmetry  to all orders in $\alpha'$ 
or, equivalently, to all orders in worldsheet loops \cite{Sen:1991zi}. 
We generalize previous one-loop results 
by computing the two-loop beta functions for bosonic string theory in cosmological backgrounds (i.e.~with target-space 
fields depending only on time), which  gives the first $\alpha'$ correction to the cosmological target-space equations.

In \cite{Bonezzi:2021mub} we computed the complete duality invariant one-loop beta function for bosonic string theory 
in a Kaluza-Klein formulation with $n$ external and $d$ internal coordinates (with target-space fields being 
independent of the internal coordinates), completing earlier results in \cite{Berman:2007xn,Berman:2007yf}. 
We showed that vanishing of the Weyl anomaly implies the two-derivative 
equations determined by  Maharana and Schwarz \cite{Maharana:1992my}. 
To this end we employed Tseytlin's  $O(d,d;\mathbb{R})$ invariant formulation  of 
the worldsheet sigma model, generalized 
to include all internal and external Kaluza-Klein fields \cite{Schwarz:1993mg,Blair:2016xnn,Bonezzi:2020ryb}.
This duality invariant sigma model is based on doubled but chiral scalars, for which there is no 
manifestly Lorentz invariant action. 
Rather, the action (in Euclidean signature) for external scalars $X^{\mu}$ and internal (doubled) scalars $Y^M$  employs the methods of Floreanini-Jackiw \cite{Floreanini:1987as} and is given by 
\begin{equation}\label{sigmaOddIntro}
\begin{split}
S[X,Y]=\frac{1}{4\pi\alpha'}\int d^2x\,\Big[&g_{\mu\nu}(X)\,\del^\alpha X^\mu\del_\alpha X^\nu+i\,\epsilon^{\alpha\beta}B_{\mu\nu}(X)\,\del_\alpha X^\mu\del_\beta X^\nu  \\
&-i\,\del_1Y^M\del_2Y_M+\cH_{MN}(X)\,\del_1Y^M\del_1Y^N\Big]\;.   
\end{split}    
\end{equation}
Here, $g_{\mu\nu}$ and $B_{\mu\nu}$ denote the external metric and B-field, respectively, and we momentarily 
suppress  the couplings to dilaton 
and Kaluza-Klein vectors. Moreover,  the ``generalized metric'' ${\cal H}_{MN}$ is an $O(d,d)$ covariant matrix encoding 
the internal metric and B-field. 
The second-order field equations for $Y^M$ can be integrated to first-order self-duality or chirality constraints, 
which can be used to show that this theory  is classically equivalent to the standard Polyakov action.

Due to the presence of chiral bosons it is to be expected that various symmetries may become anomalous, 
hence casting doubt on the quantum consistency of this model. 
Indeed, it was shown in \cite{Bonezzi:2020ryb} that duality invariance itself becomes anomalous. 
However, the anomalous transformation of the one-loop effective action can be canceled by using the Green-Schwarz mechanism, assigning a suitable transformation 
to the external B-field. This observation  gives  a worldsheet interpretation to the target-space analysis  
in \cite{Eloy:2019hnl,Eloy:2020dko} that showed that in general $O(d,d)$ transformations
need to be $\alpha'$-deformed,\footnote{This mimics an earlier observation 
in double field theory \cite{Siegel:1993th,Hull:2009mi,Hohm:2010pp}, in which the generalized gauge transformations also need to 
be $\alpha'$-deformed \cite{Hohm:2013jaa,Hohm:2014eba,Hohm:2014xsa,Marques:2015vua}, see also \cite{Bergshoeff:1995cg,Lescano:2016grn,Hohm:2016lge,Baron:2017dvb,Hronek:2020xxi,Hassler:2020wnp,Lescano:2021lup,Chang:2021tsj}.} 
although this deformation vanishes in the cosmological setting that we focus on here. 
Similarly, a theory of chiral bosons generally displays  gravitational anomalies, but one expects that for an equal number 
of left-moving and right-moving bosons, as in Tseytlin's model, 
 these anomalies cancel \cite{Sonnenschein:1988ug,Bastianelli:1990ev,Giaccari:2008zx}. 
 This was confirmed explicitly in  \cite{Bonezzi:2020ryb}
for the special case of constant backgrounds.

In this paper we revisit the quantum consistency of the duality invariant sigma model once more, motivated 
by certain complications that arise  at two loops.   
Indeed, computing the two-loop beta functions naively, by employing the minimal subtraction (MS) scheme,  
leads to results that are incompatible with duality invariance and hence are inconsistent. 
Rather than indicating an inconsistency of the model, this problem is due to the 
MS scheme not being applicable in this context, a fact that was appreciated in the 1980s in the closely related 
realm of chiral fermions \cite{Schubert:1988ke}. 
In section 2 we revisit these issues for the duality invariant Tseytlin model. We confirm in particular 
that the Lorentz symmetry does not have genuine anomalies and explain why the MS scheme cannot be 
used for renormalization. Consequently, 
the actual computation of higher-loop beta functions meets with significant technical challenges. 

In the second part of this paper we will thus change gears and compute the duality invariant beta functions 
using the standard Polyakov action restricted to cosmological backgrounds. While this action is not $O(d,d;\mathbb{R})$ invariant, the complete  beta functions can be efficiently determined  
by demanding duality invariance. This procedure will be explained in section 3 by revisiting the one-loop computation 
and then be used in section 4 to determine the ${\cal O}(\alpha')$ coefficient in the cosmological classification of 
\cite{Hohm:2019jgu}, thereby providing an independent check of \cite{Meissner:1996sa,Hohm:2015doa}, 
where this coefficient was computed by direct  dimensional reduction. 
Remarkably, the beta functions of the cosmological Polyakov action 
lead to the duality invariant beta functions (and hence field equations) rather directly, which is in contrast to the dimensional reduction procedure that requires some elaborate field redefinitions \cite{Meissner:1996sa,Hohm:2015doa,Codina:2020kvj,Codina:2021cxh}. 
At one and two loops it is easy to see diagrammatically that potential duality violating terms (that would have to be 
removed by field redefinitions) do not arise. 
At higher loops it is guaranteed that the contributions corresponding to  single-trace terms 
in the action are duality invariant, while the multi-trace terms may require field redefinitions 
in order to display duality invariance.

We close the introduction by discussing in more detail 
the subtleties of higher-loop computations in presence of chiral fields. 
Consider the quantum effective action $\Gamma[X]$ obtained by integrating out the $Y^M$ from the action \eqref{sigmaOddIntro}:  
 \begin{equation}
e^{-\Gamma[X]} = Z^{-1} \int\cD Y e^{-S[X,Y]}\;. \end{equation}
Although Lorentz invariance of the classical action (\ref{sigmaOddIntro})  is non-standard,  the $X^{\mu}$ do transform conventionally 
as scalars, so that the Lorentz symmetry of $\Gamma$, if present, must be manifest. 
The absence of genuine gravitational and hence Lorentz anomalies\footnote{A symmetric and conserved energy-momentum tensor $T_{\alpha\beta}$ yields  a conserved Lorentz current $J_\alpha=T_{\alpha\beta}\epsilon^{\beta\gamma}x_\gamma$.  Conversely, an anomalous Ward identity for $J_\alpha$ has to descend from a gravitational anomaly.} 
to be confirmed  here, however, 
does \textit{not} imply that the effective action has no anomalous transformation. It only implies that 
its anomalous transformation can be canceled by adding a finite and local counterterm \cite{Alvarez-Gaume:1983ihn}. 
In Section \ref{Sec:Quantum odd} we will explicitly show that the one-loop effective action 
receives a  Lorentz breaking contribution of the form 
\begin{equation}\label{Lorentz anomaly}
\Delta\Gamma=\int d^2x\,A_{\mu\nu}(X)\,\del_1X^\mu\del_1X^\nu\;,    
\end{equation}
with a finite and local target-space tensor $A_{\mu\nu}$. This implies that the anomalous contribution can be eliminated by adding a finite and local counterterm (for instance, we could simply take minus the term (\ref{Lorentz anomaly})),
thereby restoring Lorentz symmetry at one loop. 
 While this means that the model is consistent at the quantum level, it also implies that the minimal subtraction scheme cannot be used for renormalization \cite{Schubert:1988ke}, because here 
counterterms are fixed by demanding that the quantum effective action be finite and that the counterterms have no finite part.
Rather,  finite counterterms have to be added at every loop order to restore Lorentz invariance. This fact, in particular, makes renormalization (and thus computation of the beta function) beyond one loop particularly convoluted. The main technical difficulty is that computing the divergent part of the effective action is not enough to fix the counterterms. In order to check for Lorentz invariance, and eventually restore it, at a given loop order one has to compute also the finite part of the effective action, which is considerably more complicated than the divergent one.\footnote{There are also proposals for manifestly Lorentz-invariant formulations, see \cite{Pasti:1996vs, Pulmann:2020omk}, but it remains to be seen whether they can be employed for higher-loop computations.}

\textit{Note added:} In the process of finalizing this paper the preprint \cite{Copland:2021cyu}  appeared, which reports  on 
unsuccessful attempts to compute the two-loop beta function for the Tseytlin model in the cosmological setting. 
Their computation proceeds with a minimal subtraction scheme but, as argued here, this is 
invalid beyond one loop given the presence of harmless but non-vanishing Lorentz anomalies.

\newpage

\section{Quantum consistency}\label{Sec:Quantum odd}

We revisit the quantum consistency of the duality invariant sigma model (\ref{sigmaOddIntro}). 
In the first subsection we show the absence of gravitational anomalies by verifying the Ward identities. 
In the second subsection we compute the one-loop quantum effective action and determine in particular 
the removable Lorentz ``anomaly'' $\Delta\Gamma$ in (\ref{Lorentz anomaly}).

\subsection{Ward identities}

We study the duality invariant sigma model on a flat Euclidean worldsheet, thereby ignoring the coupling to the dilaton, and suppress the Kaluza-Klein vectors $\cA_\mu{}^M$, which play no role in the present discussion. The complete action $S[X,Y]$ splits as $S[X,Y]=S_P[X]+S_T[X,Y]$, where $S_P$ denotes the usual Polyakov sigma model for the external sector, while $S_T$ contains the fields which transform non-trivially under T-duality\footnote{Strictly speaking, at order $\alpha'$ also the external B-field transforms non-trivially under $O(d,d)$, 
due to the Green-Schwarz deformation.}: 
\begin{equation}\label{sigma Odd}
\begin{split}
S_P[X]&=\frac{1}{2\lambda}\int d^2x\,\Big[g_{\mu\nu}(X)\,\del^\alpha X^\mu\del_\alpha X^\nu+i\,\epsilon^{\alpha\beta}B_{\mu\nu}(X)\,\del_\alpha X^\mu\del_\beta X^\nu\Big]\;, \\
S_T[X,Y]&=\frac{1}{2\lambda}\int d^2x\,\Big[-i\,\del_1Y^M\del_2Y_M+\cH_{MN}(X)\,\del_1Y^M\del_1Y^N\Big]\;.   
\end{split}    
\end{equation}
In here,  $x^2$ is the Euclidean time, $\epsilon^{12}=+1$ and $\lambda=2\pi\alpha'$ is the worldsheet coupling and loop counting parameter. The worldsheet fields $X^\mu$, with $\mu=1,\ldots,D$, are the coordinate embeddings of the ``external'' directions, while $Y^M$, $M=1,\ldots,2d$, correspond to the doubled ``internal'' coordinates. The generalized metric $\cH_{MN}$ is an $O(d,d)$ matrix and thus obeys $\cH_{MP}\,\eta^{PQ}\,\cH_{QN}=\eta_{MN}$, where $\eta_{MN}$ is the $O(d,d)-$invariant metric, which is used to raise and lower fundamental $O(d,d)$ indices. It will often be convenient to work with the matrix $\cS_M{}^N:=\cH_{MP}\eta^{PN}$, which leads to simpler index-free matrix manipulations. The $O(d,d)$ constraint, for instance, translates to the simple $\cS^2=1$.  
The above action is 2D Lorentz (or rather $SO(2)$) invariant under the non-standard transformations
\begin{equation}\label{Lorentz transf}
\delta X^\mu=\xi^\alpha\del_\alpha X^\mu\;,\quad \delta Y^M=\xi^1\del_1Y^M-i\,\xi^2\del_1Y^N\cS_N{}^M \;,   
\end{equation}
where $\xi^\alpha(x)$ is a diffeomorphism which specializes to an $SO(2)$ parameter as $\xi^\alpha=\omega\,\epsilon^{\alpha\beta}x_\beta$. It should be emphasized that Lorentz invariance (at least under the transformations \eqref{Lorentz transf}) requires the constraint $\cS^2=1$. Applying the Noether procedure one finds the symmetric  and traceless energy-momentum tensor $T_{\alpha\beta}$. %=T_{\alpha\beta}^X+T_{\alpha\beta}^Y$, where
%\begin{equation}
%\begin{split}
%T_{\alpha\beta}^X&=\frac{1}{\lambda}\,\Big[g_{\mu\nu}\,\del_\alpha X^\mu\del_\beta X^\nu-\tfrac12\,\delta_{\alpha\beta}\,\big(g_{\mu\nu}\,\del X^\mu\cdot\del X^\nu\big)\Big]\;,\\ T_{11}^Y&=-T_{22}^Y=\frac{1}{2\lambda}\,\cH_{MN}\del_1Y^M\del_1Y^N\;, 
%\\ T_{12}^Y&=T_{21}^Y=-\frac{i}{2\lambda}\,\del_1Y^M\del_1Y_M\;.
%\end{split}    
%\end{equation}
The conserved Lorentz current is then given by $J_\alpha=T_{\alpha\beta}\,\epsilon^{\beta\gamma}x_\gamma$.  
As usual, it is convenient to introduce complex coordinates\footnote{The standard notation for the complex coordinates is $(z,\bar z)$, but we find it more convenient to use the $x^\pm$ notation in order to keep track of the $SO(2)$ tensor character.} on the plane as $x^\pm=x^1\pm ix^2$, yielding $\del_1=\del_++\del_-$ and $\del_2=i(\del_+-\del_-)$. The two independent components of $T_{\alpha\beta}$ can similarly be rewritten as
\begin{equation}\label{T}
\begin{split}
T_{++}&=T_{11}-iT_{12}=\frac{2}{\lambda}\,g_{\mu\nu}(X)\,\del_+ X^\mu\del_+ X^\nu-\frac{1}{2\lambda}\,\Big(\eta_{MN}-\cH_{MN}(X)\Big)\,\del_1Y^M\del_1Y^N\;,\\
T_{--}&=T_{11}+iT_{12}=\frac{2}{\lambda}\,g_{\mu\nu}(X)\,\del_- X^\mu\del_- X^\nu+\frac{1}{2\lambda}\,\Big(\eta_{MN}+\cH_{MN}(X)\Big)\,\del_1Y^M\del_1Y^N\;,
\end{split}    
\end{equation}
and obey the holomorphic conservation law $\del_- T_{++}=0$, $\del_+ T_{--}=0$.
Since the Lorentz symmetry is not manifest, it can in principle become anomalous  at the quantum level. Given that the energy-momentum tensor is manifestly symmetric, the conservation of the Lorentz current is guaranteed by the conservation of the energy-momentum tensor. If $T_{\alpha\beta}$ is not conserved at the quantum level, then there is  a gravitational anomaly in curved space, which renders the theory inconsistent.

After these preliminaries we now prove the absence of gravitational anomalies. 
In two dimensions, anomalies arise in two-point functions of conserved currents, such as $\langle T(x)T(y)\rangle$. In order to compute the two-point functions of the energy-momentum tensor \eqref{T} in the nonlinear sigma model, one has to expand the target-space fields around a fixed spacetime point, such as
\begin{equation}\label{expansion}
g_{\mu\nu}(X)=\eta_{\mu\nu}+\cO(X)\;,\quad  \cH_{MN}(X)=H_{MN}+\cO(X)\;,   
\end{equation}
where $H_{MN}$ is a \emph{constant} $O(d,d)$ matrix obeying $H_{MP}\,\eta^{PQ}\,H_{QN}=\eta_{MN}$. As for the full $\cH_{MN}(X)$, we define $S_M{}^N=H_{MP}\,\eta^{PN}$, which obeys $S^2=1$. Since every contribution of order $X$ and higher in the expansion \eqref{expansion} generates a trilinear and higher order vertex in the energy momentum tensor \eqref{T}, they only contribute to the $\langle TT\rangle$ two-point function starting at two loops. Given that anomalies only arise at one-loop level, we are left with the two-point functions of the energy-momentum tensor of the free theory:
\begin{equation}\label{free T}
\begin{split}
T_{++}&=\frac{2}{\lambda}\,\del_+ X^\mu\del_+ X_\mu-\frac{1}{\lambda}\,\Pi^-_{MN}\,\del_1Y^M\del_1Y^N\;,\\ 
 T_{--}&=\frac{2}{\lambda}\,\del_- X^\mu\del_- X_\mu+\frac{1}{\lambda}\,\Pi^+_{MN}\,\del_1Y^M\del_1Y^N\;,
\end{split}
\end{equation}
for which the $X$-part and the $Y$-part, in the following denoted by $T_{\alpha\beta}^X$ and $T_{\alpha\beta}^Y$, respectively,  are separately conserved.
In \eqref{free T}  we have introduced the constant $O(d,d)$ projectors $\Pi^\pm_{MN}=\frac12\,\big(\eta_{MN}\pm H_{MN}\big)$, whose matrices $\Pi_\pm=\frac12\,\big(1\pm S\big)$ satisfy $\Pi_\pm^2=\Pi_\pm$ as well as $\Pi_\pm\Pi_\mp=0$.
The constant backgrounds $\eta_{\mu\nu}$ and $H_{MN}$ also define the propagators of the free theory, which read
\begin{equation}\label{propagators}
\begin{split}
\langle X^\mu(x)X^\nu(y)\rangle&=\lambda\,\eta^{\mu\nu}\int\frac{d^2p}{(2\pi)^2}\frac{e^{ip\cdot (x-y)}}{p^2}\;,\\
\langle Y^M(x)Y^N(y)\rangle&=\lambda\,\int\frac{d^2p}{(2\pi)^2}\frac{e^{ip\cdot (x-y)}}{p^2}\,\Big[H^{MN}+i\,\frac{p_2}{p_1}\,\eta^{MN}\Big]\;.
\end{split}
\end{equation}
Since at this level there are no interactions between $X$ and $Y$, the connected two-point functions split as sums of the form
\begin{equation}
\l T(x)T(y)\r=\l T^X(x)T^X(y)\r+\l T^Y(x)T^Y(y)\r \;,   
\end{equation}
which obey the Ward identities separately.
Since the Polyakov part of the sigma model cannot generate gravitational anomalies, one has only to compute the connected two point functions of $T_{\alpha\beta}^Y$, and we will drop the superscript $Y$ from now on. Since the matrix $S$ obeys $S\Pi_\pm=\Pi_\pm S=\pm\Pi_\pm$, it is easy to see that the mixed correlator $\l T_{++}(x)T_{--}(y)\r$ vanishes due to the $O(d,d)$ tensor structure. Upon performing the Wick contractions using the propagators \eqref{propagators} one finds
\begin{equation}
\begin{split}
\l T_{++}(x) T_{++}(y)\r&=\int\frac{d^2p}{(2\pi)^2}\,e^{ip\cdot(x-y)}\l T_{++}T_{++}\r(p)\;,\\
\l T_{--}(x) T_{--}(y)\r&=\int\frac{d^2p}{(2\pi)^2}\,e^{ip\cdot(x-y)}\l T_{--} T_{--}\r(p)
\end{split}    
\end{equation}
with the momentum space expressions 
\begin{equation}\label{TT}
\begin{split}
\l T_{++}T_{++}\r(p)&=\frac{d}{2}\,\Big[J_1(p)-2i\,J_2(p)-J_3(p)\Big]\;,\\
\l T_{--}T_{--}\r(p)&=\frac{d}{2}\,\Big[J_1(p)+2i\,J_2(p)-J_3(p)\Big]\;,
\end{split}    
\end{equation}
where we used ${\rm Tr}\Pi_\pm=d$, while the one-loop integrals read
\begin{equation}\label{J defined}
\begin{split}
J_1(p)&=\int\frac{d^2k}{(2\pi)^2}\,\frac{k_1^2(p_1+k_1)^2}{k^2(p+k)^2}\;,\\ 
J_2(p)&=\int\frac{d^2k}{(2\pi)^2}\,\frac{k_1k_2(p_1+k_1)^2}{k^2(p+k)^2}\,, \\
J_3(p)&=\int\frac{d^2k}{(2\pi)^2}\,\frac{k_1k_2(p_1+k_1)(p_2+k_2)}{k^2(p+k)^2}\;.
\end{split}    
\end{equation}
The separate integrals $J_i(p)$ are UV quadratically divergent. We thus employ dimensional regularization by extending the integrations as
\begin{equation}
\int\frac{d^{2}k}{(2\pi)^{2}}\;\rightarrow\;\mu^{2-n}\int\frac{d^{n}k}{(2\pi)^{n}}  \;,\quad n=2+\epsilon\;,  
\end{equation}
thereby introducing the renormalization scale $\mu$.
The components $k_1$ and $k_2$ entering explicitly in \eqref{J defined} are treated as two particular components of the $n-$dimensional vector $k_\alpha$. The integrals are then computed by using the manifest $SO(n)$ symmetry of the denominators and integration measure, yielding
\begin{equation}\label{J123}
\begin{split}
J_1(p)&=\frac{1}{4\pi}\,\left[p^2\,\left(\frac{1}{4\epsilon}-\frac13+\frac18\,\log\frac{p^2}{\mu^2}\right)+\frac{p_1^2}{6}+\frac{p_1^4}{6p^2}\right]\;,\\ J_2(p)&=\frac{1}{4\pi}\,\left[\frac{p_1p_2}{12}+\frac{p_1^3p_2}{6p^2}\right] \;,\\
J_3(p)&=\frac{1}{4\pi}\,\left[p^2\,\left(\frac{1}{4\epsilon}-\frac14+\frac18\,\log\frac{p^2}{\mu^2}\right)+\frac{p_1^2p_2^2}{6p^2}\right]\;,
\end{split}    
\end{equation}
where we have redefined $4\pi e^{-\gamma}\mu^2\rightarrow\mu^2$, with $\gamma$ the Euler-Mascheroni constant. 
Summing the contributions to find the correlators \eqref{TT} one can see that both the UV divergence and the renormalization scale disappear, leaving
\begin{equation}
\begin{split}
\l T_{++} T_{++}\r(p)&=\frac{d}{8\pi}\,\Big[-\frac{p^2}{12}+\frac{p_1^2}{6}+\frac{p_1^2(p_1^2-p_2^2)}{6p^2}-\frac{i\,p_1p_2}{6}-\frac{i\,p_1^3p_2}{3p^2}\Big]\;,\\
\l T_{--} T_{--}\r(p)&=\frac{d}{8\pi}\,\Big[-\frac{p^2}{12}+\frac{p_1^2}{6}+\frac{p_1^2(p_1^2-p_2^2)}{6p^2}+\frac{i\,p_1p_2}{6}+\frac{i\,p_1^3p_2}{3p^2}\Big]\;.
\end{split}    
\end{equation}
The result can be rewritten in terms of complex momenta $p_\pm=\frac12\,(p_1\pm ip_2)$, giving the more familiar form
\begin{equation}\label{correlator}
\begin{split}
\l T_{++} T_{++}\r(p)&=\frac{d}{48\pi}\,\left[\frac{p_+^3}{p_-}+4\,p_+^2+p_+p_-\right]\;,\\
\l T_{--} T_{--}\r(p)&=\frac{d}{48\pi}\,\left[\frac{p_-^3}{p_+}+4\,p_-^2+p_+p_-\right]\;.
\end{split}    
\end{equation}
One can see that the non-local part of the correlator coincides with the one obtained from standard scalars in two dimensions (as, for instance, the $X^\mu$). This part gives a non-vanishing gravitational anomaly that, as usual, can be canceled at the expense of the Weyl symmetry, \emph{i.e.}~by generating an anomalous trace of the energy-momentum tensor $T_{+-}$. While the extra local terms in the correlator \eqref{correlator} cannot give rise to genuine anomalies\footnote{By coupling the theory to 2D gravity, the $\l TT\r$ two point functions give the order $h^2$ contribution to the effective action of the metric fluctuation $h_{\alpha\beta}$. Any unwanted local term can be removed by adding a suitable local counterterm to the original action.}, one can see that the local contributions violate the rigid $SO(2)$ symmetry in flat space. Indeed, $\l T_{++}T_{++}\r$ is a ``spin'' $+4$ tensor, which is respected by the non-local contribution $\frac{p_+^3}{p_-}=\frac{4\,p_+^4}{p^2}$, but the local terms in \eqref{correlator} have spin $+2$ and zero.

This analysis implies that the sigma model cannot have genuine gravitational anomalies. This ensures that the effective action constructed from \eqref{sigma Odd} can be made Lorentz invariant at the quantum level. The result \eqref{correlator}, however, suggests that the model does have trivial anomalies, that have to be canceled by adding suitable local counterterms to the action \eqref{sigma Odd}. We will substantiate this claim in the next section by computing the one-loop effective action in a perturbative expansion in fields.

\subsection{Effective action}

At this point, one could study the effective action of the sigma model \eqref{sigma Odd} by employing the background field method \cite{Abbott:1981ke,AlvarezGaume:1981hn}. However, rather than studying the full one-loop effective action, we shall focus on the partial effective action generated by integrating out the $Y$ fields from $S[X,Y]$: 
\begin{equation}
e^{-\Gamma[X]}=\frac{1}{Z}\,\int\cD Y\,e^{-S[X,Y]}\;. 
\end{equation}
We shall stress that $\Gamma[X]$ encodes \emph{only} the quantum effects of $Y$-loops on the dynamics of the $X$-sector. Since the $X^\mu$ fields are normal scalars under 2D Lorentz transformations, this is sufficient to study the Lorentz symmetry of the $X$-sector at the quantum level. Indeed, the advantage in studying the effective action $\Gamma[X]$ is that its Lorentz invariance (or breaking thereof) is manifest.
 Upon splitting $S[X,Y]$ into the Polyakov and Tseytlin parts as in \eqref{sigma Odd}, one can see that the effective action $\Gamma$ is given by the sum of the classical Polyakov action $S_P$ and of the one-loop correction $W$ due to the $Y$ path integral. The latter is given by
\begin{equation}
e^{-W[X]}=\frac{1}{Z}\int\cD Y\,e^{-S_T[X,Y]}\;,
\end{equation}
where we recall that the Tseytlin action reads
\begin{equation}\label{Tseytlin action}
S_T[X,Y]=\frac{1}{2\lambda}\int d^2x\,\Big[-i\,\del_1Y^M\del_2Y_M+\cH_{MN}(X)\,\del_1Y^M\del_1Y^N\Big]\;.    
\end{equation}
In order to extract a standard kinetic term for the $Y^M$, one usually flattens the fields
by means of generalized vielbeins $E_M{}^A$ and defining $Y^M=E^M{}_A(X)\,Y^A$, which generates further interactions with the Maurer-Cartan form $\Omega_\mu^{AB}=E_M{}^A\del_\mu E^{MB}$. This way of proceeding, however, seems inconvenient due to the infrared divergence of the $Y-$propagator \eqref{propagators} for $p_1\to0$, which is removed only if all $Y$'s appear under a $\del_1$. At the price of loosing manifest background independence, we shall thus introduce a \emph{constant} background $H_{MN}$, and expand the full generalized metric in powers of fluctuations. This can be achieved by considering the linearized fluctuation $h_{MN}$ as the fundamental field. By using the matrix $\cS$ and defining again $S$ as the matrix obtained by raising one index of the background $H_{MN}$, the expansion can be written as $\cS=S+h+\cO(h^2)$. Since both the full $\cS$ and the background $S$ square to one, we choose a power series ansatz for the expansion, where one verifies the constraint $\cS^2=1$ order by order in $h$. It is easy to see that all odd powers of $h$ besides the linear one can be removed by $O(d,d)$ covariant field redefinitions. Using an ansatz containing only even powers of $h$ one can solve the constraint in closed form by the formula
\begin{equation}
\cS=S+h+S\Delta(h)\;,\quad \Delta(h)=\sqrt{1-h^2}-1=-\tfrac12\,h^2+\cO(h^4)\;.    
\end{equation}
At this point, the leading term $H_{MN}\,\del_1Y^M\del_1Y^N$ in the expansion of \eqref{Tseytlin action} is part of the free kinetic term giving the propagator \eqref{propagators}, while the rest yields the interacting part
\begin{equation}\label{Sint}
V_h=\frac{1}{2\lambda}\int d^2x\,\Big\{\big(h_{MN}+(S\Delta)_{MN}\big)\,\del_1Y^M\del_1Y^N\Big\}\;.  
\end{equation}
The interaction $V_h$ can be used to determine $W[X]$ as a power series in $h$ which gives, to second order,
\begin{equation}\label{W}
W[X]=-\big\l e^{-V_{h}}\big\r_{\rm 1PI}=\big\l V_h\big\r_{\rm 1PI}-\tfrac12\,\big\l V_h^2\big\r_{\rm 1PI}+\cO(h^3)\;.   
\end{equation}
The term $\big\l V_h\big\r$ is a tadpole, whose Feynman integral reads
\begin{equation}
\int\dk\frac{k_1^2}{k^2}=\frac{1}{n}\int\dk\,1=0\;,    
\end{equation}
vanishing in dimensional regularization\footnote{We shall give more details on the computation of Feynman integrals in the next sections.}. We shall thus evaluate the bubble diagram $\big\l V_h^2\big\r_{\rm 1PI}$ to lowest order in $h$, which yields
\begin{equation}
-\frac12\,\big\l V_h^2\big\r_{\rm 1PI}=\frac14\,\int\frac{d^2p}{(2\pi)^2}\,\Pi(p)\,\tr\big(h(p)\,h(-p)\big) \;,
\end{equation}
where $\Pi$ is given by $\Pi(p)=J_1(p)+J_3(p)$ in terms of the integrals \eqref{J123}:
\begin{equation}
\Pi(p)=\frac{1}{4\pi}\,\left\{p^2\,\left(\frac{1}{2\epsilon}+\frac14\,\log\frac{p^2}{\mu^2}-\frac{7}{12}\right)+\frac13\,p_1^2\right\}\;.    
\end{equation}
The order $h^2$ effective action thus splits as $W=W_{\rm n.l.}+W_{\rm div.}+W_{\rm loc.}+\cO(h^3)$. The divergent term $W_{\rm div.}$ is local and Lorentz invariant:
\begin{equation}
W_{\rm div.}=\frac{1}{32\pi\epsilon}\int d^2x\,\tr\big(\nabla_\mu h\nabla_\nu h\big)\,\del^\alpha X^\mu\del_\alpha X^\nu\;.
\end{equation}
The non-local contribution $W_{\rm n.l.}$ is Lorentz invariant as well and is given in momentum space by
\begin{equation}
W_{\rm n.l.}=\frac{1}{64\pi}\int\frac{d^2p}{(2\pi)^2}\,\log\frac{p^2}{\mu^2}\,\tr\Big[\big(\del^\alpha X^\mu\nabla_\mu h\big)(p)\,\big(\del_\alpha X^\nu\nabla_\nu h\big)(-p)\Big]\;.
\end{equation}
The absence of genuine gravitational anomalies ensures that the divergent and the non-local parts of the effective action are Lorentz invariant, which we have confirmed by this direct computation. The local part of the effective action, on the other hand, has two contributions: $W_{\rm loc.}=W_{\rm inv.}+\Delta\Gamma$, where $W_{\rm inv.}=-\frac{7}{192\pi}\int d^2x\,\tr\big(\nabla_\mu h\nabla_\nu h\big)\,\del^\alpha X^\mu\del_\alpha X^\nu$ is Lorentz invariant, while $\Delta\Gamma$ displays a trivial anomaly, which is given by
\begin{equation}\label{Delta Gamma}
\Delta\Gamma=\frac{1}{48\pi}\int d^2x\,\tr\Big[\nabla_\mu h\nabla_\nu h\Big]\,\del_1 X^\mu\del_1 X^\nu\;.
\end{equation}
Since $\Delta\Gamma$ is finite and local, Lorentz invariance can be restored at this order by adding a counterterm, the simplest choice being $\Delta S_{\rm Lorentz}=-\Delta\Gamma$.

Let us comment on this result. The appearance of Lorentz violating terms in the effective action $\Gamma[X]$ is expected in dimensional regularization, since the theory contains chiral bosons and the Lorentz symmetry is not manifest. This does not imply an inconsistency, since the Lorentz violating terms are local and can be eliminated by a finite local counterterm. The renormalized one-loop effective action is thus finite and Lorentz invariant.
The necessity of adding these finite counterterms, however, prevents one from using the minimal subtraction scheme for renormalization. Focusing on the beta functions, this issue does not affect their computation at one-loop level, but makes higher-loop computations very complicated.
In order to illustrate these complications, let us recall the standard string sigma model with metric coupling $g_{\mu\nu}$, where counterterms can be extracted from the so-called bare action, defined in $2+\epsilon$ dimensions as $S_0=\frac{1}{2\lambda}\int d^{2+\epsilon}x\,g_{0\mu\nu} \del^\alpha X^\mu\del_\alpha X^\nu$. The bare metric $g_{0\mu\nu}$ is then related to the renormalized metric $g_{\mu\nu}$ by
\begin{equation}
g_{0\mu\nu}=\mu^\epsilon\Big[g_{\mu\nu}+\sum_{L=1}^\infty\sum_{n=1}^L\frac{1}{\epsilon^n}\,T_{\mu\nu}^{(L,n)}(g)\Big]  \;,  
\end{equation}
where the (purely divergent) counterterms are parametrized by the tensors $T_{\mu\nu}^{(L,n)}$ and are fixed by demanding that they cancel the divergences arising from loop diagrams. Knowing the counterterms, the beta function $\beta_{\mu\nu}=\frac{d g_{\mu\nu}}{d\log\mu}$ is determined by imposing that the bare coupling $g_{0\mu\nu}$ does not depend on the renormalization scale $\mu$. In contrast, due to the contribution \eqref{Delta Gamma}, for the Tseytlin model considered here 
a single bare metric cannot contain all needed counterterms. In order to proceed with standard renormalization one is forced to define the bare action as
\begin{equation}
S_0=\frac{1}{2\lambda}\int d^{2+\epsilon}x\,\Big[g^{\scalebox{0.55}{(1)}}_{0\mu\nu}\,\del_1 X^\mu\del_1 X^\nu+g^{\scalebox{0.55}{(2)}}_{0\mu\nu}\,\del_2 X^\mu\del_2 X^\nu\Big] \;,   
\end{equation}
in terms of two independent bare couplings $g^{\scalebox{0.55}{(1,2)}}_{0\mu\nu}$. An analogous procedure was employed, for instance, in \cite{Schubert:1988ke} to renormalize four-dimensional theories with chiral fermions. These independent couplings are related to the single renormalized coupling $g_{\mu\nu}$ (otherwise Lorentz symmetry would be explicitly broken) 
via both divergent and finite counterterms as
\begin{equation}
g^{\scalebox{0.55}{(i)}}_{0\mu\nu}=\mu^\epsilon\Big[g_{\mu\nu}+\sum_{L=1}^\infty\sum_{n=0}^L\frac{1}{\epsilon^n}\,T_{\mu\nu}^{{\scalebox{0.55}{(i)}}(L,n)}\Big]  \;.  
\end{equation}
The finite counterterms, in particular, have to restore Lorentz invariance by imposing that the finite quantum corrections to $g_{\mu\nu}$ coincide in the two sectors $\del_1X^\mu\del_1X^\nu$ and $\del_2X^\mu\del_2X^\nu$, thereby defining a single quantum effective coupling that multiplies $\del^\alpha X^\mu\del_\alpha X^\nu$.
For the case at hand, our explicit computation determines the one-loop counterterms to be\footnote{We omit the standard Polyakov counterterms involving, for instance, the Ricci tensor $R_{\mu\nu}$.}
\begin{equation}
\begin{split}
g^{\scalebox{0.55}{(1)}}_{0\mu\nu}&=\mu^\epsilon\Big[g_{\mu\nu}-\frac{\lambda}{16\pi\epsilon}\,\tr\big(\nabla_\mu h\nabla_\nu h\big)-\frac{\lambda}{24\pi}\,\tr\big(\nabla_\mu h\nabla_\nu h\big)\Big]\;,\\
g^{\scalebox{0.55}{(2)}}_{0\mu\nu}&=\mu^\epsilon\Big[g_{\mu\nu}-\frac{\lambda}{16\pi\epsilon}\,\tr\big(\nabla_\mu h\nabla_\nu h\big)\Big]\;,
\end{split}    
\end{equation}
which yield a finite and Lorentz invariant effective action in terms of the single renormalized metric $g_{\mu\nu}$. Thanks to the absence of true Lorentz anomalies, this must be the case at any loop order. The main technical difficulty, however, is that in order to determine the counterterms (and thus the beta function) it is not sufficient to compute the divergent part of the effective action. In order to restore Lorentz invariance, at any loop level one has to compute the finite part as well, which is considerably more complicated. 
%Due to this technical impediment we shall abandon for the moment the duality invariant sigma model. In the next sections we will %thus study the standard Polyakov sigma model on cosmological backgrounds. We will show that, upon using a suitable %perturbative expansion, the corresponding beta functions can be used to derive $O(d,d)$ invariant field equations with relative %minimum effort.

\section{Cosmological Polyakov 
at one loop}

In this section we propose an alternative  method, circumventing  the above mentioned technical complications, 
 to determine the $\alpha'$ corrections of the target-space effective action in the simplest cosmological reduction
analyzed in \cite{Hohm:2019jgu}. 
We use the standard Polyakov sigma model but applied directly to cosmological backgrounds and  show that, by using a suitable worldsheet perturbative expansion, 
$O(d,d)$ invariant  beta functions  can be straightforwardly recovered.

\subsection{Classical action}

We start by considering a $(d+1)-$dimensional space\footnote{We choose a target space with Euclidean signature in order to have a positive definite worldsheet Euclidean action.} where we split coordinates as $x^\mu=(t,y^i)$ and make the ansatz that target-space fields only depend on (Euclidean) time $t$. In order to maximally simplify the computation, we will further set the B-field $B_{ij}(t)$ and the mixed components of the metric $g_{ti}(t)$ to zero. The metric and dilaton take then the form
\begin{equation}\label{cosmo ansatz}
g_{\mu\nu}=
\begin{pmatrix}
n^2(t) & 0 \\
0 & g_{ij}(t)
\end{pmatrix} \;,\quad \Phi=\Phi(t)\;,   
\end{equation}
which breaks the original $\mathfrak{diff}_{d+1}$ symmetry down to $\mathfrak{diff}_1\times GL(d)$. Having set the B-field to zero, the duality group is reduced to $GL(d)\times\mathbb{Z}_2$, but we will assume that the full $O(d,d)$ group would be restored by switching on $B_{ij}(t)$.

Using this ansatz in the Polyakov sigma model gives the worldsheet action
\begin{equation}\label{cosmo sigma}
S=\frac{1}{2\lambda}\int d^2x\,\Big[n^2(t)\,\del^\alpha t\,\del_\alpha t+g_{ij}(t)\,\del^\alpha Y^i\del_\alpha Y^j\Big]\;,
\end{equation}
where $\lambda=2\pi\alpha'$ is the loop-counting parameter and $x^\alpha$ are two-dimensional Euclidean coordinates. We choose, as usual, to work on a flat worldsheet, thereby discarding the dilaton coupling. As  is well known \cite{Tseytlin:1986ws,Curci:1986hi,Bonezzi:2021mub}, the dilaton contribution to the field equations is fixed on general grounds by the Weyl anomaly equations, once the beta functions of the other couplings are known. At this point, one may think that it would be more convenient to work with a single worldsheet coupling $g_{\mu\nu}(X)$ and use the ansatz \eqref{cosmo ansatz} at the end of the computation. It turns out, instead, that considering $n^2(t)$ as a one-dimensional metric and $g_{ij}(t)$ as a $GL(d)$ multiplet of scalars from the beginning is crucial in order to produce manifest duality invariant field equations.
In order to determine the beta functions of the sigma model \eqref{cosmo sigma} in the minimal subtraction scheme, one needs to compute the divergent part of the quantum effective action, which we will do in the following. 

%\subsection{Background field expansion}

As a preliminary for the computation of the beta functions we must now perform a background field expansion, 
which is a 
convenient way to compute the effective action. One proceeds by shifting the fields as $t\rightarrow t+\pi$ and $Y^i\rightarrow Y^i+\pi^i$, where now $t$ and $Y^i$ are viewed as classical backgrounds, and integrate over the quantum fluctuations $\pi$ and $\pi^i$. Since we want to preserve manifest $\mathfrak{diff}_1$ covariance, we shall redefine $\pi=\pi(\xi)$ in terms of a covariant fluctuation $\xi$, which is a genuine one-dimensional vector.
This allows us to expand the shifted action in a simple way \cite{Mukhi:1985vy,Howe:1986vm}:
\begin{equation}\label{cosmo expansion}
S[t+\pi(\xi), Y+\pi]=e^{\xi\cD}  S[t, Y+\pi]\;,  
\end{equation}
where $\cD=n^{-1}\del_t$ is the one-dimensional covariant derivative\footnote{The vector fluctuation $\xi$ is taken with a flat index: $\pi(\xi)=n^{-1}\xi+\cO(\xi^2)$, and the covariant derivative $\cD$ corresponds to $\nabla_a=e^\mu_a\,\nabla_\mu$ in higher dimensions.}, with the additional rules
\begin{equation}\label{rules}
\xi\cD\,\xi=0\;,\quad \xi\cD\,\del_\alpha t=n^{-1}\del_\alpha\xi\;,\quad[\xi\cD, \del_\alpha]=0\;,   \end{equation}
when acting on worldsheet fields. In order to compute the $L$-loop effective action one needs to expand the shifted action up to order $2L$ in fluctuations. In particular, this means that at one loop one needs only the terms quadratic in fluctuations from the expansion \eqref{cosmo expansion}, which are given by
\begin{equation}\label{cosmo S2}
\begin{split}
S_2&=\frac{1}{2\lambda}\int d^2x\,\Big[\del^\alpha\xi\del_\alpha\xi+g_{ij}\,\del^\alpha\pi^i\del_\alpha\pi^j+2\,\xi\,\cD g_{ij}\,\del^\alpha Y^i\del_\alpha\pi^j+\tfrac12\,\xi^2\,\cD^2g_{ij}\,\del^\alpha Y^i\del_\alpha Y^j\Big]\, ,    
\end{split}    
\end{equation}
where all target-space fields are evaluated at $t$. Since the kinetic term of $\pi^i$ involves $g_{ij}(t)$, which depends on $x^\alpha$, it is useful to introduce internal vielbeins $e_i{}^a(t)$ and the flat $SO(d)$ metric $\delta_{ab}$ to flatten the fluctuation as $\pi^i=e^i{}_a\,\pi^a$. This introduces a local $SO(d)$ symmetry realizing the coset $GL(d)/SO(d)$, where the symmetric tensor $g_{ij}=e_i{}^ae_j{}^b\delta_{ab}$ is viewed as a standard representative. The worldsheet derivatives indeed covariantize as
\begin{equation}
\del_\alpha\pi^i=e_a{}^i\,D_\alpha\pi^a\;,\quad D_\alpha\pi^a=\del_\alpha\pi^a+n\,\del_\alpha t\,W^a{}_b\,\pi^b\;,\quad W^{ab}=e_i{}^a\cD e^{bi}\;.
\end{equation}
The Maurer-Cartan connection $W^{ab}$ splits into an antisymmetric component $Q^{ab}=W^{[ab]}$, which is a genuine connection for local $SO(d)$ rotations, plus a symmetric tensor $P^{ab}=W^{(ab)}$.
The action \eqref{cosmo S2} splits as $S_2=S_{\rm kin}+V_2+V_{\rm mix}$ into a kinetic term plus interaction vertices, given by
\begin{equation}
\begin{split}
S_{\rm kin}&=\frac{1}{2\lambda}\int d^2x\,\Big[\del^\alpha\xi\del_\alpha\xi+\del^\alpha\pi^a\del_\alpha\pi_a\Big]\;,\\
V_2&=\frac{1}{2\lambda}\int d^2x\,\Big[2n\,\del^\alpha t\,W_{ab}\,\del_\alpha\pi^a\pi^b+n^2\del^\alpha t\del_\alpha t\,W^c{}_a\,W_{cb}\,\pi^a\pi^b\\
&\hspace{23mm}+2\,\xi\,\cD g_{ij}e_a{}^j\del^\alpha Y^i\del_\alpha\pi^a+\tfrac12\,\xi^2\,\cD^2g_{ij}\,\del^\alpha Y^i\del_\alpha Y^j\Big]\;.
\end{split}    
\end{equation}
The extra term $V_{\rm mix}$, which depends on the connection $W^{ab}$, is proportional to $\del_\alpha t\del^\alpha Y^i$ and thus cannot contribute to UV divergences.

\subsection{One-loop beta function}

The full one-loop effective action is given by the normalized\footnote{We choose the normalization such that $\l1\r=1$.} expectation value
\begin{equation}
\Gamma_{1l}=-\big\l e^{-(V_2+V_{\rm mix})}\big\r_{\rm 1PI}\;,    
\end{equation}
where the subscript 1PI implies to keep only one-particle irreducible contributions.
Wick contractions are performed with the two-point functions
\begin{equation}
\l\xi(x)\xi(y)\r=\lambda \int\frac{d^2p}{(2\pi)^2}\,\frac{e^{ip\cdot x}}{p^2}\;,\quad \l\pi^a(x)\pi^b(y)\r=\lambda\,\delta^{ab} \int\frac{d^2p}{(2\pi)^2}\,\frac{e^{ip\cdot x}}{p^2}\;.   
\end{equation}
Renormalizability of the sigma model \eqref{cosmo sigma} ensures that the only divergent parts of $\Gamma$ are local and proportional to either $\del^\alpha t\del_\alpha t$ or $\del^\alpha Y^i\del_\alpha Y^j$. For this reason, at one-loop order we only need to compute
\begin{equation}
\Gamma_{1l\,{\rm div}}\subset\big\l V_2\big\r-\tfrac12\,\big\l V_2^2\big\r_{\rm 1PI}   \;. 
\end{equation}
The resulting Feynman integrals contain both ultraviolet (UV) and infrared (IR) divergences. We use standard dimensional regularization by extending every loop integration to $n=2+\epsilon$ dimensions and substituting
\begin{equation}
\int\frac{d^2k}{(2\pi)^2}\longrightarrow\tilde\mu^{2-n}\int\dk \;,\quad n=2+\epsilon\;, \end{equation}
where $\tilde\mu$ is the renormalization scale. IR divergences can be regularized in various ways. One option is to substitute every massless propagator with $\frac{1}{p^2+m^2}$. This, however, requires to add a mass term to the Lagrangian that has to be renormalized and used in the background field expansion, which makes higher-loop computations harder. We instead choose to regularize Feynman integrals by putting masses only in those propagators which actually cause infrared divergences at zero external momenta. The basic one-loop integral requiring both UV and IR regularization is the tadpole\footnote{In higher dimensions massless tadpoles are zero in dimensional regularization, but not in two dimensions.} $\int\frac{d^2k}{(2\pi)^2}\,\frac{1}{k^2}$, which is regularized as
\begin{equation}\label{I}
\begin{split}
I&=\tilde\mu^{2-n}\int\dk\,\frac{1}{k^2+m^2}=\frac{1}{4\pi}\,\left(\frac{m^2}{4\pi\tilde\mu^2}\right)^{\epsilon/2}\,\Gamma(-\epsilon/2)\\
&=-\frac{1}{2\pi}\,\left(\frac{1}{\epsilon}+\frac12\,\log\frac{m^2}{\mu^2}\right)+\cO(\epsilon)\;,
\end{split}
\end{equation}
where $\mu^2=4\pi\,e^{-\gamma}\tilde\mu^2$. We will also make frequent use of the identity
\begin{equation}
\int\dk\,1=0\;,    
\end{equation}
which can be derived by computing
\begin{equation}
\tilde\mu^{2-n}\int\dk\frac{k^2}{k^2+m^2}=-m^2I\;.    
\end{equation}
With these rules in place one can easily compute the tadpole diagrams $\l V_2\r$, yielding
\begin{equation}\label{tadpole}
\l V_2\r=\frac{I}{4}\int d^2x\,\cD^2g_{ij}\,\del^\alpha Y^i\del_\alpha Y^j+\frac{I}{2}\int d^2x\,n^2\,W^{ab}W_{ab}\,\del^\alpha t\del_\alpha t\;.
\end{equation}
The bubble diagrams contained in $\l V_2^2\r$ are generally non-local, being of the schematic form
\begin{equation}\label{Fourier bubble}
\int\frac{d^2p}{(2\pi)^2}\,A(p)\,\Pi(p)\,B(-p) \;,   
\end{equation}
where $A$ and $B$ represent the Fourier transform of products of fields, \emph{e.g.} $\del_\alpha t\,n(t)\,W^{ab}(t)$, while $\Pi(p)$ is the one-loop  bubble integral. The only UV divergent bubbles at one loop are given by
\begin{equation}
\Pi_{\alpha\beta}(p)=\int\frac{d^2k}{(2\pi)^2}\,\frac{k_\alpha k_\beta}{k^2(p-k)^2} \;,\quad \Pi'_{\alpha\beta}(p)=\int\frac{d^2k}{(2\pi)^2}\,\frac{k_{(\alpha}(p-k)_{\beta)}}{k^2(p-k)^2}\;,  
\end{equation}
and one can see by power counting that their UV divergence only comes from the zero-momentum contribution $\Pi(0)=\Pi'(0)$, which gives the local expression
\begin{equation}
\Pi(0)\int d^2x\,A(x)\,B(x)\;.
\end{equation}
The divergent part $\Pi(0)$ is regularized as
\begin{equation}
\begin{split}
\Pi_{\alpha\beta}(0)&=\tilde\mu^{2-n}\int\dk\,\frac{k_\alpha k_\beta}{k^2(k^2+m^2)}=\frac{1}{n}\,\delta_{\alpha\beta}\,\tilde\mu^{2-n}\int\dk\,\frac{k^2}{k^2(k^2+m^2)}\\
&=\frac{1}{n}\,\delta_{\alpha\beta}\,I=-\frac{1}{4\pi\epsilon}\,\delta_{\alpha\beta}+\cO(\epsilon^0)\;.    
\end{split}
\end{equation}
With these techniques we can compute the divergent part of $-\tfrac12\,\big\l V_2^2\big\r_{\rm 1PI}$, which reads
\begin{equation}\label{bubble}
\begin{split}
-\tfrac12\,\big\l V_2^2\big\r_{\rm 1PI}=&-\frac{I}{4}\int d^2x\,\big(\cD gg^{-1}\cD g\big)_{ij}\,\del^\alpha Y^i\del_\alpha Y^j\\
&-\frac{I}{4}\int d^2x\,n^2\,W^{ab}\big(W_{ab}-W_{ba}\big)\del^\alpha t\del_\alpha t\;.  \end{split}    
\end{equation}
Summing the contributions of \eqref{tadpole} and \eqref{bubble} one obtains the full one-loop divergences. These have to be canceled, in the MS scheme, by purely divergent counterterms.  In the $Y-$sector we obtain
\begin{equation}\label{Y 1loop count}
S_{Y\,{\rm c.t.}}=\frac{1}{8\pi\epsilon}\int d^2x\,\big(\cD^2g-\cD gg^{-1}\cD g\big)_{ij}\,\del^\alpha Y^i\del_\alpha Y^j \;,   
\end{equation}
while the $t-$sector gives
\begin{equation}\label{T 1loop count}
\begin{split}
S_{t\,{\rm c.t.}}&=\frac{1}{8\pi\epsilon}\int d^2x\,n^2\,W^{ab}\big(W_{ab}+W_{ba}\big)\,\del^\alpha t\del_\alpha t\\
&=\frac{1}{4\pi\epsilon}\int d^2x\,n^2P^{ab}P_{ab}\,\del^\alpha t\del_\alpha t\\
&=-\frac{1}{16\pi\epsilon}\int d^2x\,n^2\,\tr\big(\cD g\cD g^{-1}\big)\,\del^\alpha t\del_\alpha t\;.
\end{split}    
\end{equation}
Comparing a generic one-loop counterterm of the form
\begin{equation}
S_{c.t.}=\frac{1}{\epsilon}\int d^2x\,\Big[\cM\,\del^\alpha t\del_\alpha t+\cM_{ij}\,\del^\alpha Y^i\del_\alpha Y^j\Big]\;,    
\end{equation}
with the original action \eqref{cosmo sigma}, one can read off the form of the bare couplings $n^2_0$ and $g_{0ij}$ as
\begin{equation}
n^2_0=\mu^\epsilon\left(n^2+\frac{2\lambda}{\epsilon}\,\cM\right)\;,\quad g_{0ij}=\mu^\epsilon\left(g_{ij}+\frac{2\lambda}{\epsilon}\,\cM_{ij}\right)\;.
\end{equation}
This in turn fixes the one-loop beta functions (see \cite{Bonezzi:2021mub}, for instance, for details) to be
\begin{equation}\label{one-loop betas}
\beta=-2\lambda\,\cM\;,\quad\beta_{ij}=-2\lambda\,\cM_{ij}\;.    
\end{equation}
Applying \eqref{one-loop betas} to the counterterms \eqref{Y 1loop count} and \eqref{T 1loop count} and recalling that $\lambda=2\pi\alpha'$, we finally obtain
\begin{equation}
\beta=\frac{\alpha'}{4}\,n^2\,\tr\big(\cD g\cD g^{-1}\big)\;,\quad \beta_{ij}=-\frac{\alpha'}{2}\,\big(\cD^2g-\cD gg^{-1}\cD g\big)_{ij}\;.    
\end{equation}

\subsection{Duality invariance}\label{Sec:duality}

Let us discuss the duality covariance of these beta functions. In the general case where the internal B-field $B_{ij}(t)$ is non vanishing, $g_{ij}$ and $B_{ij}$ can be combined into the manifestly $O(d,d)$ covariant generalized metric $\cS_M{}^N$, which we write in matrix form as
\begin{equation}\label{full S}
\cS=\begin{pmatrix}
Bg^{-1} & g-Bg^{-1}B\\
g^{-1} & -g^{-1}B
\end{pmatrix}   \;, \quad\cS^2=1\;, 
\end{equation}
where we treat $g$, $B$ and $g^{-1}$ as $GL(d)$ matrices. When $B_{ij}$ is zero, the generalized metric simplifies to
\begin{equation}\label{simple S}
\cS_M{}^N=\begin{pmatrix}
0&g_{ij}\\g^{ij}&0
\end{pmatrix}\;,
\end{equation}
which is covariant only under $GL(d)\times\mathbb{Z}_2$. It is now quite simple to see that strings of $GL(d)$ matrix products of the form
\begin{equation}
\big(\cD^{p_1}gg^{-1}\cD^{p_2}gg^{-1}\cdots g^{-1}\cD^{p_n}g\big)_{ij}    
\end{equation}
coincide with the ${}_{ij}$ component of the duality covariant tensor
\begin{equation}
\big(\cD^{p_1}\cS\,\cS\cD^{p_2}\cS\,\cS\cdots \cS\cD^{p_n}\cS\big)_M{}^N \;,   
\end{equation}
in terms of the simplified $\cS$ in \eqref{simple S}. Traces require more care: in general, $GL(d)$ traces will not combine into $O(d,d)$ ones prior to computing the Feynman integrals. At this level, however, it is simple to see that the $GL(d)$ trace $\tr\big(\cD g\cD g^{-1}\big)$ can be written in terms of an $O(d,d)$ trace as
\begin{equation}
\tr\big(\cD g\cD g^{-1}\big)=\tfrac12\,\Tr\big(\cD\cS\big)^2\big|_{B=0}\;.
\end{equation}
In a more general setting with an arbitrary number of external dimensions, one does not expect the beta function to be manifestly $O(d,d)$ covariant, due to the Green-Schwarz deformation \cite{Eloy:2019hnl,Eloy:2020dko}. In the cosmological setting, however, this deformation vanishes being a target-space two-form. 

We have thus shown that the one-loop beta functions can be written in the form
\begin{equation}\label{1l betas}
\beta=\frac{\alpha'}{8}\,n^2\,\Tr\big(\cD\cS\big)^2\;,\quad \beta_{ij}=-\frac{\alpha'}{2}\,\big(\cD^2\cS-\cD\cS\cS\cD\cS\big)_{ij}\;.    
\end{equation}
This is not enough to prove covariance of $\beta_{ij}$ under the $\mathbb{Z}_2$ $T-$duality: given a generic covariant tensor
\begin{equation}
\cM_M{}^N=\begin{pmatrix}
0&M_{ij}\\\widetilde M^{ij}&0
\end{pmatrix} \;,   
\end{equation}
the $\mathbb{Z}_2$ duality acts by swapping $M_{ij}\leftrightarrow\widetilde M^{ij}$, which is induced by the $\mathbb{Z}_2$ operation $g\leftrightarrow g^{-1}$. Given only $M_{ij}$, however, there is no simple way to construct the dual $\widetilde M^{ij}$ with $GL(d)$ operations.
In order to do so, it is useful to introduce a basis of tensors with definite parity under conjugation by $\cS$. For instance, from $\cS^2=1$ one deduces that $\cD\cS$ has odd parity, meaning that
\begin{equation}\label{fundamental oddness}
\cS\big(\cD\cS\big)\cS=-\cD\cS\;.    
\end{equation}
Higher derivatives of $\cS$, however, do not have definite parity under conjugation. Let us define the $O(d,d)$ matrix
\begin{equation}\label{T tensor}
\cT=\cD^2\cS+\cS\big(\cD\cS\big)^2 \;,  \quad \cT_M{}^N=\begin{pmatrix}
0&T_{ij}\\\widetilde T^{ij}&0
\end{pmatrix} \;,
\end{equation}
so that $\beta_{ij}=-\frac{\alpha'}{2}\,T_{ij}$. Contrary to $\cD^2\cS$, the matrix $\cT$ has definite parity: 
\begin{equation}\label{STS}
\cS\cT\cS=-\cT \;,   
\end{equation}
as can be seen by taking a derivative of \eqref{fundamental oddness}. More generally, 
given an arbitrary $O(d,d)$ matrix $\cM$, one can decompose it into parity eigenstates by projection:
\begin{equation}\label{projection}
\cM=\cM_++\cM_-\;,\quad \cM_\pm=\tfrac12\,\Big(\cM\pm\cS\cM\cS\Big)    \;,
\end{equation}
with the projected components obeying $\cS\cM_\pm\cS=\pm\cM_\pm$.
When the B-field is zero, $\cM_M{}^N$ has only components $M_{ij}$ and $\widetilde M^{ij}$. Upon decomposing $\cM$ into $\cM_+$ and $\cM_-$, one can see that the projected components are simply related by conjugation by $g$, namely
\begin{equation}
M_{\pm\,ij}=\pm g_{ik}\,\widetilde M^{kl}_\pm\,g_{lj}\;.
\end{equation} 
Since the tensor $\cT$ has odd parity, we see that the component $\beta_{ij}$ is sufficient in order to define $\boldsymbol{\beta}_M{}^N$ by
\begin{equation}
\boldsymbol{\beta}_M{}^N=\begin{pmatrix}
0&\beta_{ij}\\\widetilde\beta^{ij}&0
\end{pmatrix}\;,\quad\widetilde\beta^{ij}=-g^{ik}\beta_{kl}g^{lj}\;.    
\end{equation}

Having computed the beta functions, the target-space equations are given by the vanishing of the anomaly functions $\bar\beta$ and $\bar\beta_{ij}$ \cite{Tseytlin:1986ws,Curci:1986hi,Hull:1987yi,Bonezzi:2021mub}. Specializing the formulas of \cite{Bonezzi:2021mub} to the cosmological case one has
\begin{equation}\label{beta bar}
\begin{split}
\bar\beta&=\beta+2\alpha'n^2\,\cD^2\Phi+n^2\,\cD\cW\;,\\
\bar\beta_{ij}&=\beta_{ij}+\alpha'\,\cD\Phi\cD g_{ij}+\tfrac12\,\cW\cD g_{ij}\;,
\end{split} 
\end{equation}
where $\Phi$ is the dilaton and $\cW$ is a one-dimensional vector constructed from $g_{ij}$ and $\cD$. One can compute $\cW$ by renormalization of the operators $\del^\alpha t\del_\alpha t$ and $\del^\alpha Y^i\del_\alpha Y^j$ \cite{Tseytlin:1986ws,Curci:1986hi,Metsaev:1987zx}, but it is more convenient to determine it by other means. The vector $\cW$ has an expansion in $\alpha'$ of the form $\cW=\sum_{n=1}^\infty{\alpha'}^{\,n}\,\cW_n$, where $\cW_n$ contains $2n-1$ derivatives $\cD$. At lowest order, one-dimensional covariance and rigid $GL(d)$ invariance fix its form up to a constant:
\begin{equation}
\cW_1=k_1\,\tr\big(g^{-1}\cD g\big)=k_1\,\cD\big(\log\det g\big)\;. \end{equation}
In order to fix the constant $k_1$, we \emph{demand} duality covariance of the target-space equations. To this end, let us notice that $\Phi$ is the standard dilaton, which does transform under $T-$duality. The duality-invariant dilaton $\phi$ is given by
\begin{equation}
\phi=2\,\Phi-\tfrac12\,\log\det g\;,
\end{equation}
and one can see that by taking $k_1=-\frac12$ all terms not writable in terms of $\cS$ or $\phi$ cancel, yielding the field equations
\begin{equation}\label{eoms Z2}
\begin{split}
\frac18\,n^2\,\Tr\big(\cD\cS\big)^2+n^2\,\cD^2\phi&=0\\
E_{ij}=\Big(\cT-\cD\phi\cD\cS\Big)_{ij}&=0\;.
\end{split}    
\end{equation}
The scalar equation is manifestly $\mathbb{Z}_2$ invariant. The $GL(d)$ tensor equation $E_{ij}=0$ instead  requires the dual equation $\widetilde E^{ij}=0$ to be satisfied. Since $\cT$ and $\cD\cS$ are both parity odd, this is indeed the case, given that $\widetilde E^{ij}=-g^{ik}E_{kl}g^{lj}$. We have thus shown that the equations \eqref{eoms Z2} are duality invariant and we shall write them in covariant matrix form:
\begin{equation}\label{eoms Odd 1l}
\Tr\big(\cD\cS\big)^2+8\,\cD^2\phi=0\;,\quad
\cD^2\cS+\cS\big(\cD\cS\big)^2-\cD\phi\cD\cS=0\;. \end{equation}
As a further consistency check, one should notice that a field equation for $\cS$ must be parity odd in order to descend from an action principle. This is due to $\delta\cS$ being odd, and it can easily seen to be the case for \eqref{eoms Odd 1l}.
The target-space equations must be $O(d,d)$ invariant once $B_{ij}$ is turned on. Given that the equations \eqref{eoms Odd 1l} are manifestly covariant, we invoke $O(d,d)$ symmetry to extend \eqref{eoms Odd 1l} to the case with non-vanishing $B_{ij}$, where now $\cS$ is the full generalized metric \eqref{full S}.
The field equations \eqref{eoms Odd 1l} coincide with the ones of the cosmological low-energy effective action \cite{Meissner:1996sa,Hohm:2019jgu}:
\begin{equation}\label{target 0}
I_0=\int dt\,n\,e^{-\phi}\Big[-\tfrac18\,\Tr\big(\cD\cS\big)^2-\big(\cD\phi\big)^2\Big]    \;.
\end{equation}
Let us notice that the action of \cite{Hohm:2019jgu} is written for a Lorentzian target-space metric. The mapping to our Euclidean metric is given, at the worldsheet level, by $n^2\rightarrow-n^2$. Given the same two-derivative action\footnote{Upon going to Einstein frame the Euclidean Lagrangian is positive definite.} \eqref{target 0}, the coefficients of the ${\alpha'}^{\,n}$ corrections will have a relative $(-1)^n$ sign compared to the notation of \cite{Hohm:2019jgu}.

In order to determine the dilaton equation, one demands mutual consistency of the other field equations \cite{Bonezzi:2021mub}. In this case, one applies a covariant derivative $\cD$ to the first equation in \eqref{eoms Odd 1l}. Upon substituting the on-shell value of $\cD^2\cS$ and $\Tr\big(\cD\cS\big)^2$, as well as using the identity
\begin{equation}\label{dilaton eom 1l}
\Tr\Big(\cS\big(\cD\cS\big)^n\Big)=0\;,\quad n\geq0\;,
\end{equation}
one finds $\cD(\cD^2\phi-\cD\phi\cD\phi)=0$. The three derivative equation can be integrated, with the integration constant being proportional to $\frac{d-25}{\alpha'}$ \cite{Tseytlin:1986ws}. The constant term vanishes in the critical dimension $d+1=26$, yielding
\begin{equation}
\cD^2\phi-\big(\cD\phi\big)^2=0\;,    
\end{equation}
which is equivalent to the dilaton equation derived from \eqref{target 0}.

\section{Cosmological Polyakov at two loops}

In this section we turn to our second main result: the computation of the duality invariant  two-loop beta function 
and the  determination of the order $\alpha'$ correction to the cosmological target-space equations. 
We begin in the first subsection with a discussion of the ambiguities of beta functions, which reflect the 
ambiguities in the target-space theory due to field redefinitions. 
In the second subsection we compute the two-loop beta function, which will then be used 
in the third subsection to determine the target-space theory. We close, in the final subsection,  with a discussion of possible simplifications for higher-loop computations.

\subsection{Beta function ambiguities}
Before starting the computation of the two-loop beta functions, we shall discuss a strategy to simplify the problem. The low-energy spacetime effective action has an expansion in powers of $\alpha'$ of the form $I=\sum_{n=0}^\infty{\alpha'}^{\,n} I_n$, where the two-derivative action $I_0$ is given by \eqref{target 0}. The classification of \cite{Hohm:2019jgu} implies that the first order correction $I_1$ is determined, modulo field redefinitions, by a single parameter $c_2$ as
\begin{equation}\label{target 1}
I_1=c_2\int dt\,n\,e^{-\phi}\,\Tr\big(\cD\cS\big)^4  \;. 
\end{equation}
From the sigma model perspective, this implies that it is sufficient to determine the field equation of $\cS$ in order to fix $c_2$. We will thus compute the two-loop beta function only for the $g_{ij}$ coupling. 

Let us suppose to have computed the beta function $\beta_{ij}$ at two loops in the MS scheme. On general grounds, one expects it to be  $GL(d)$ covariant. Furthermore, as we mentioned in section \ref{Sec:duality}, our perturbative expansion ensures that every term without traces can be written in terms of the duality covariant matrix $\cS$. This is not the case for terms containing traces, where $O(d,d)$ covariance can in general be established only after performing the Feynman integrals and possibly 
$GL(d)$-covariant  field redefinitions. It turns out that at two loops the only $O(d,d)$-breaking trace term is $\tr\big(g^{-1}\cD^2g\big)$ but, as we will show in the following, its  coefficient is zero. This establishes that $\beta_{ij}$ can be written in terms of $\cS$ at two loops, without the need for any field redefinitions. 

Coming now to possible ambiguities arising from field redefinitions, we shall assume that
$\beta_{ij}(g,\cD)=\beta_{ij}(\cS,\cD)$, for $B=0$.  Expanding in $\alpha'$ one has
\begin{equation}\label{beta expansion}
\begin{split}
\beta_{ij}&=\alpha'\beta_{ij}^{\,\scalebox{0.55}{(1)}}+{\alpha'}^{\,2}\beta_{ij}^{\,\scalebox{0.55}{(2)}}+\cO\big({\alpha'}^{\,3}\big)\;,\\
\beta&=\alpha'\beta^{\,\scalebox{0.55}{(1)}}+\cO\big({\alpha'}^{\,2}\big)\;,
\end{split}
\end{equation}
where $\alpha'\beta_{ij}^{\,\scalebox{0.55}{(1)}}$ and $\alpha'\beta^{\,\scalebox{0.55}{(1)}}$ are given by \eqref{1l betas}.
The beta functions are computed in a given renormalization scheme, which we choose to be minimal subtraction. A change in the renormalization scheme is equivalent to a redefinition of the couplings $n^2$ and $g_{ij}$ \cite{Metsaev:1987zx}. While this does not affect the one-loop beta functions, it introduces an ambiguity starting at two loops. Given a set of couplings $\varphi^I=(n^2,g_{ij})$, they can be viewed as a set of coordinates in the (infinite dimensional) coupling space. Since the beta functions are given by $\beta^I=\frac{d\varphi^I}{d\log\mu}$, they are tangent vectors along renormalization group trajectories. A field redefinition of $n^2$ and $g_{ij}$ can be viewed as a change of coordinates in coupling space: $\varphi^I\rightarrow\varphi^I+\delta\varphi^I$, under which $\beta^I$ transforms as a vector, \emph{i.e.} $\delta\beta^I=\cL_{\delta\varphi}\beta^I$, where $\cL$ denotes the Lie derivative. For the case at hand this results in
\begin{equation}\label{ambiguity general}
\delta\beta_{ij}=\delta g_{kl}\cdot\frac{\del\beta_{ij}}{\del g_{kl}}+\delta n^2\cdot\frac{\del\beta_{ij}}{\del n^2}-\beta_{kl}\cdot\frac{\del(\delta g_{ij})}{\del g_{kl}}-\beta\cdot\frac{\del(\delta g_{ij})}{\del n^2}\;,   
\end{equation}
where derivatives are functional derivatives acting as
\begin{equation}
\begin{split}
f_{ij}\cdot\frac{\del A[g]}{\del g_{ij}}&=\int dt\,f_{ij}(t)\,\frac{\delta}{\delta g_{ij}(t)}\,A[g] =A[g+f]\rvert_{\text{linear part in}\,f}\;,\\
h\cdot\frac{\del A[n^2]}{\del n^2}&=\int dt\,h(t)\,\frac{\delta}{\delta n^2(t)}\,A[n^2] =A[n^2+h]\rvert_{\text{linear part in}\,h}\;.
\end{split}   
\end{equation}
Since the spacetime action \eqref{target 1} is given in a fixed field basis, it is necessary to account for the ambiguity \eqref{ambiguity general} to be able to compare equations of motion.
Given that $\beta_{ij}^{\,\scalebox{0.55}{(2)}}$ can already be written in terms of $\cS$, we only look at field redefinitions for which $\delta\beta_{ij}$ can also be written in terms of $\cS$. The most general redefinition with this property is given, at order $\alpha'$, by
\begin{equation}\label{field redef}
\begin{split}
\delta n^2&=a_1\,\alpha'n^2\,\Tr\big(\cD\cS\big)^2\;,\\
\delta g_{ij}&=\alpha'\,\Big[b_1\,\cT+b_2\,\cS\big(\cD\cS\big)^2\Big]_{ij}\;.
\end{split}    
\end{equation}
Using \eqref{field redef} in \eqref{ambiguity general} one obtains (apart from $\delta\beta_{ij}^{\,\scalebox{0.55}{(1)}}=0$)
\begin{equation}\label{ambiguity}
\delta\beta_{ij}^{\,\scalebox{0.55}{(2)}}=
p\,\Big(\cT\,\Tr\big(\cD\cS\big)^2+\cD\cS\, \Tr\big(\cT\cD\cS\big)\Big)_{ij}+q\,\Big(\tfrac18\,\cS\big(\cD\cS\big)^2\Tr\big(\cD\cS\big)^2-\cS\cT^2\Big)_{ij}\;,
\end{equation}
where $p=\frac12\,a_1+\frac18\,b_1$ and $q=b_2$. The remaining ingredient to write down the field equations $\bar\beta_{ij}=0$ at two loops is the $\cW$ vector appearing in \eqref{beta bar}. At order ${\alpha'}^{\,2}$, $\cW_2$ is a $GL(d)$ scalar with three derivatives $\cD$ acting on $g_{ij}$ and its inverse. There are several $GL(d)$ invariant possibilities, but requiring $O(d,d)$ invariance fixes $\cW_2$ to be of the form
\begin{equation}\label{W2}
\cW_2=k_2\,\Tr\big(\cT\cD\cS\big)\;.
\end{equation}
Having collected these ingredients, the target-space equations can be written as
\begin{equation}\label{eom 2loop general}
-\tfrac12\,\big(\cT-\cD\phi\cD\cS\big)_{ij}+\alpha'\,\big(\beta_{ij}^{\,\scalebox{0.55}{(2)}}+\delta\beta_{ij}^{\,\scalebox{0.55}{(2)}}+\tfrac12\,\cW_2\cD g_{ij}\big) =0\;,   
\end{equation}
which should be compared with the equations obtained from
the action $I_0+\alpha'I_1$.
As we have discussed in the previous section, any variational equation for $\cS$ obtained from an action $I$ must be of definite odd parity. This will provide a useful check of our computation, since any parity even term arising from $\beta_{ij}^{\,\scalebox{0.55}{(2)}}$ should be removable by a suitable choice of parameters in $\delta\beta_{ij}^{\,\scalebox{0.55}{(2)}}$. 
Having discussed how to relate the two-loop beta function with field equations, in the next section we will compute $\beta_{ij}^{\,\scalebox{0.55}{(2)}}$ by determining the divergent part of the two-loop effective action.

\subsection{Effective action and two-loop beta function}

The full effective action $\Gamma$ has a meaningful expansion in powers of $\del_\alpha Y^i$ as
\begin{equation}\label{full Gamma}
\Gamma=\Gamma_{0\del Y}+\Gamma_{1\del Y}+\Gamma_{2\del Y}+\cdots\;.    
\end{equation}
UV divergent terms are present only in $\Gamma_{0\del Y}$, which determines the beta function for $n^2$, and in $\Gamma_{2\del Y}$, determining the beta function for $g_{ij}$. We then restrict to $\Gamma_{2\del Y}$ and compute its UV divergences. Renormalizability implies that the divergent part of $\Gamma_{2\del Y}$ contains no factors of $\del_\alpha t$. We shall thus focus on the smaller subsector $\Gamma_{2\del Y,0\del t}$, which we name $\Gamma_g$.  
The relevant vertices at two loops can be obtained by applying the background field expansion on the $Y-$sector of \eqref{cosmo sigma}, namely
\begin{equation}
S_Y=\frac{1}{2\lambda}\int d^2x\,g_{ij}(t)\,\del^\alpha Y^i\del_\alpha Y^j\;,
\end{equation}
up to fourth order in fluctuations. The resulting interaction part, $S_{Y\, \text{int}}$, can be decomposed in terms of the number of external legs $\del_\alpha Y^i$ as follows:
\begin{equation}
\begin{split}
S_{Y\, \text{int}} &= V_{0Y} + V_{1Y} + V_{2Y}\;,\\
V_{0Y} &=\frac{1}{2 \lambda} \ZeroVa{D}{L}{D} +  \frac{1}{4 \lambda} \ZeroVb{D}{L}{L}{D}\\
V_{1Y} &=\frac{1}{\lambda} \OneVa{D}{L} + \frac{1}{2 \lambda} \OneVb{D}{L}{L} + \frac{1}{6 \lambda} \OneVc{D}{L}{L}{L}\;, \\
V_{2Y} &=\frac{1}{4 \lambda} \TwoVa{L}{L} + \frac{1}{12 \lambda} \TwoVb{L}{L}{L} + \frac{1}{48 \lambda} \TwoVc{L}{L}{L}{L}\;,
\end{split}    
\end{equation}
where we choose the following representation for the vertices
\begin{equation}
\begin{split}
\ZeroVa{D}{L}{D} &= \int d^2 x\, \xi\,\cD g_{ij}e_a{}^i e_b{}^j \del^\alpha \pi^a\del_\alpha\pi^b\;,\qquad\ZeroVb{D}{L}{L}{D} = \int d^2 x\, \xi^2\,\cD^2 g_{ij}e_a{}^i e_b{}^j \del^\alpha \pi^a\del_\alpha\pi^b\;,\\[2mm]
\OneVa{D}{L} &= \int d^2 x\, \xi\,\cD g_{ij}e_a{}^i \del^\alpha \pi^a\del_\alpha Y^j\;,\qquad
\OneVb{D}{L}{L} = \int d^2 x\, \xi^2\,\cD^2 g_{ij}e_a{}^i \del^\alpha \pi^a\del_\alpha Y^j\;,\\[2mm]
\OneVc{D}{L}{L}{L} &= \int d^2 x\, \xi^3\,\cD^3 g_{ij}e_a{}^i \del^\alpha \pi^a\del_\alpha Y^j\;,\quad\hspace{1mm}
\TwoVa{L}{L} = \int d^2 x\, \xi^2\,\cD^2 g_{ij} \del^\alpha Y^i\del_\alpha Y^j\;,\\[2mm]
\TwoVb{L}{L}{L} &= \int d^2 x\, \xi^3\,\cD^3 g_{ij} \del^\alpha Y^i\del_\alpha Y^j\;,\qquad\hspace{2mm}
\TwoVc{L}{L}{L}{L} = \int d^2 x\, \xi^4\,\cD^4 g_{ij} \del^\alpha Y^i\del_\alpha Y^j\;.
\end{split}
\end{equation}
In this diagrammatic representation, straight red lines and dashed blue lines correspond to $\xi$ and $\partial_\alpha \pi^a$ fluctuations, respectively, while wavy blue lines represent external legs $\partial_\alpha Y^i$. The tensor structure of each vertex can be read from the diagram, since each red line corresponds to one derivative of $g_{i j}$ and each dashed blue line represents an internal vielbein $e_a{}^i$. For instance, a vertex with $p$ internal red lines and one internal blue line encodes the structure $\cD^p g_{i j} e_a{}^i$. 

With these vertices, the relevant two-loop effective action is given by
\begin{equation}
\Gamma_{g,2l}=\Big\l e^{-V_{0Y}}\big(V_{2Y} - \tfrac{1}{2}\, V_{1Y}^2\big)\Big\r_{\lambda,\rm 1PI}\, -\; \text{subtractions} \;,
\end{equation}
where the subscript means to keep only contributions of order $\lambda$, which correspond to two-loop diagrams.
The role of subtraction terms is to remove one-loop subdivergences of two-loop diagrams, ensuring that all non-local divergences cancel. At two loops, the subtractions can be obtained by expanding the one-loop counterterms up to second order in fluctuations and using  the new vertices to insert counterterms in one-loop diagrams. This procedure, however, does not seem valid at higher-loop order \cite{Howe:1986vm}. In view of future applications, we shall employ a different method, which consists in subtracting subdivergences diagram by diagram \cite{Foakes:1987gg, Jack:1989vp,Kleinert:2001ax}. This also allows to show that entire classes of diagrams can be ignored when computing the beta function. We denote the subtraction procedure by an operator $\R$ acting on a given two-loop diagram. We also assume that all finite contributions are discarded at the end.
The divergent part of $\Gamma_{g,2l}$ is  then given by the sum of twelve diagrams as follows:
\begin{equation}\label{Gamma diagrams}
\begin{aligned}
\Gamma_{g,2l\, \text{div}} = \lambda\, \R &\left[ \frac{1}{16}\, \TwoLoopA{L}{L}{L}{L} - \frac{1}{2}\,\TwoLoopB{L}{L}{L}{D} - \frac{1}{8}\, \TwoLoopC{D}{D}{L}{L} + \frac{1}{4}\, \TwoLoopD{L}{L}{D}{L} \right.\\
& + \frac{1}{4}\, \TwoLoopD{D}{D}{L}{D} + \frac{1}{2}\, \TwoLoopE{L}{D}{L}{D} -\frac{1}{4}\, \TwoLoopF{L}{L}{D} + \TwoLoopG{D}{L}{L}\\
& \left. + \frac{1}{8}\, \TwoLoopH{D}{L}{D} -\frac{1}{4}\, \TwoLoopI{D}{L}{D}{L}{D} -\frac{1}{2}\, \TwoLoopI{D}{D}{L}{D}{L} -\frac{1}{2}\, \TwoLoopJ{L}{D}{L}{D}{D}\right]\;.
\end{aligned}
\end{equation}
A greatly simplifying feature of $\Gamma_{g}$ is that its divergent part, at any loop order, can be computed from diagrams with zero external momenta. In particular, this means that the diagrams in \eqref{Gamma diagrams} are in fact vacuum diagrams, with the external lines and white circles only denoting vertices.
For the sake of compactness,  the Feynman diagrams in \eqref{Gamma diagrams} represent both the Feynman integral and the worldsheet structure, \emph{e.g.}
\begin{equation}\label{prototipe}
\begin{split}
\TwoLoopB{L}{L}{L}{D}&=J_{\alpha\beta}\int d^2x\,\big(\cD^3\cS\cS\cD\cS\big)_{ij}\,\del^\alpha Y^i\del^\beta Y^j\;,\\
J_{\alpha\beta}&=\int dkdl\,\frac{k_\alpha k_\beta}{k^4l^2}\;,  
\end{split}    
\end{equation}
where we introduced a shorthand notation for the dimensionally extended measure:
\begin{equation}
\int dk=\tilde\mu^{2-n}\int\frac{d^nk}{(2\pi)^n}\;,\quad \mu^2=4\pi e^{-\gamma}\tilde\mu^2\;. \end{equation}
Before discussing the method, let us mention that all diagrams in \eqref{Gamma diagrams} fall into two topological classes. The first six diagrams belong to the ``chain'' topology, which consists of two one-loop diagrams joined at a vertex. These are the simplest to compute since the two one-loop factors do not have common momenta. The remaining six diagrams belong instead to the so-called ``sunset'' topology, where the two individual loops share momentum along a common line.

We will now present the strategy to compute the two-loop diagrams. In order to better illustrate the procedure, we will include some detailed examples before giving the final result. 
As we have previously mentioned, the Feynman integrals corresponding to \eqref{Gamma diagrams} are plagued by infrared divergences. In computing (the UV divergent part of) two-loop integrals, we will proceed as follows: 
\begin{enumerate}
\item Write down each Feynman diagram with purely massless propagators.\\[-4mm]
\item Using algebraic and integration by parts identities, manipulate numerators to rewrite every \emph{integrand} in terms of a basis of master integrals.\\[-4mm]
\item In this latter basis, put mass regulators only on propagators responsible for IR divergences.\\[-4mm]
\item Compute only the master integrals.
\end{enumerate}
It should be stressed that this four-step procedure is valid at any loop order and at higher loops simplifies the computation enormously \cite{Grisaru:1986kw,Foakes:1987gg,Jack:1989vp, Ketov:1990hw}.
From now on, the Feynman diagrams will represent only the integrals so that, for instance, the diagram \eqref{prototipe} only stands for $J_{\alpha\beta}$. At two loops, it turns out that all twelve diagrams can be reduced to linear combinations of just two master diagrams, one for each topology.

Let us start discussing the first three steps of the above list.
The first diagram in \eqref{Gamma diagrams} cannot be simplified, meaning that it is the master integral for the chain topology:
\begin{equation}\label{master1}
\TwoLoopA{L}{L}{L}{L}\;=\left(\int dk\,\frac{1}{k^2}\right)^2\;\xrightarrow{\rm IR\; reg.}\;\left(\int dk\,\frac{1}{k^2+m^2}\right)^2\;.
\end{equation}
In the second step we have introduced masses as IR regulators where necessary. The second diagram in \eqref{Gamma diagrams} can be reduced to \eqref{master1} by using integration by parts (IBP) identities in momentum space. In particular, for the 2-loop case we only need
\begin{equation}\label{IBP1}
\int dk\, \frac{\partial}{\partial k_{\alpha}} \left( \frac{k_{\beta}}{k^2}\right) = 0 \ ,
\end{equation}
yet analogous identities exist for higher number of propagators and external momenta contributions. Using \eqref{IBP1} we get
\begin{equation}\label{reduction2}
\begin{split}
\TwoLoopB{L}{L}{L}{D}&=\int dkdl\,\frac{k_\alpha k_\beta}{k^4l^2} = \left(\int dl\, \frac{1}{l^2}\right) \left[- \frac{1}{2} \int dk\, \frac{\partial}{\partial k^\alpha}\left(\frac{1}{k^2}\right) k_\beta \right]\\
&= \left(\int dl\, \frac{1}{l^2}\right) \left(\frac{1}{2} \delta_{\alpha \beta} \int dk\, \frac{1}{k^2} \right)\\
&= \frac{1}{2}\, \delta_{\alpha \beta}\, \TwoLoopA{L}{L}{L}{L} \; .
\end{split}    
\end{equation}

The third diagram in \eqref{Gamma diagrams} is zero due to the identity $\int dk\,1=0$, which arises in the diagram from the blue tadpole. The next diagram can be also reduced to \eqref{master1} by an intermediate (almost trivial) algebraic step
\begin{equation}
\begin{split}
\TwoLoopD{L}{L}{D}{L}&=\int dkdl\,\frac{k_\alpha k_\beta k^2}{k^6l^2}=\int dkdl\,\frac{k_\alpha k_\beta}{k^4l^2}=\TwoLoopB{L}{L}{L}{D} = \frac{1}{2}\, \delta_{\alpha \beta}\, \TwoLoopA{L}{L}{L}{L} \;,   
\end{split}
\end{equation}
where we used \eqref{reduction2} in the last equality. One can appreciate that starting with massless propagators is crucial for the reduction.
Apart from another diagram vanishing due to a blue tadpole, the only remaining chain diagram can be reduced to \eqref{master1} by using \eqref{IBP1} twice:
\begin{equation}\label{master3}
\TwoLoopE{L}{D}{L}{D}=\int dkdl\,\frac{k_\alpha k^\gamma}{k^4}\,\frac{l_\gamma l_\beta}{l^4} = \left(\frac{1}{2} \delta_{\alpha}{}^{\gamma} \int dk\, \frac{1}{k^2} \right) \left(\frac{1}{2} \delta_{\gamma \beta} \int dl\, \frac{1}{l^2} \right) = \frac{1}{4}\, \delta_{\alpha \beta}\,  \TwoLoopA{L}{L}{L}{L} \;. 
\end{equation}

Starting with the sunset topology, the first diagram of this type is the only sunset master integral and is given by
\begin{equation}\label{master4}
\TwoLoopF{L}{L}{D}=\int dkdl\,\frac{l_\alpha l_\beta}{k^2l^2(k-l)^2}\;\xrightarrow{\rm IR\;reg.}\;\int dkdl\,\frac{l_\alpha l_\beta}{l^2(k^2+m^2)\big((k-l)^2+m^2\big)} \;. 
\end{equation}
It is easy to see that the massive propagators correspond to the red lines in the diagram. 
Let us now give an explicit example of a reduction involving a nontrivial algebraic manipulation. We consider the next sunset diagram in \eqref{Gamma diagrams}, which is given by
\begin{equation}\label{reduction example}
\TwoLoopG{D}{L}{L}=\int dkdl\,\frac{k_{(\alpha} l_{\beta)} k\cdot l}{k^4l^2(k-l)^2} \;. 
\end{equation}
In order to proceed, we use the identity $k\cdot l=\tfrac12\,\big(k^2+l^2-(k-l)^2\big)$ to rewrite \eqref{reduction example} as
\begin{equation}
\begin{split}
\TwoLoopG{D}{L}{L}&=\frac12\int dkdl\,\left[\frac{k_{(\alpha} l_{\beta)}}{k^2l^2(k-l)^2}+\frac{k_{(\alpha} l_{\beta)}}{k^4(k-l)^2}-\frac{k_{(\alpha} l_{\beta)}}{k^4l^2}\right]\\
&=\frac12\int dkdl\,\left[\frac{k_{(\alpha} l_{\beta)}}{k^2l^2(k-l)^2}+\frac{k_{\alpha} k_{\beta}}{k^4l^2}-2\,\frac{k_{(\alpha} l_{\beta)}}{k^4l^2}\right]\;,
\end{split}
\end{equation}
where, in going from the first to the second line, we wrote $l_\beta=k_\beta-(k-l)_\beta$ and then renamed $k-l\to l$. This rewriting plus renaming trick corresponds to integrating by parts in configuration space (which has nothing to do with the IBP identity in momentum space \eqref{IBP1}).
By counting propagators one can see that the first term is of the form of \eqref{master4}, albeit with a different positioning of derivatives. The second term coincides with \eqref{reduction2}, while the last term vanishes by $SO(n)$ symmetry. At this point, we integrate by parts (in configuration space) the first term, yielding
\begin{equation}
\begin{split}
\frac12\int dkdl\,\frac{k_{(\alpha} l_{\beta)}}{k^2l^2(k-l)^2}&=\frac12\int dkdl\,\frac{k_{\alpha} k_{\beta}-k_{(\alpha} (k-l)_{\beta)}}{k^2l^2(k-l)^2}\\
&=\frac14\int dkdl\,\frac{k_{\alpha} k_{\beta}}{k^2l^2(k-l)^2}=\frac14\;\TwoLoopF{L}{L}{D}\;,
\end{split}    
\end{equation}
where we recognized the left-hand side in the second term of the first line. Putting the two terms together and using \eqref{reduction2}, we finally obtain the diagrammatic reduction
\begin{equation}
\TwoLoopG{D}{L}{L}=\frac14\;\TwoLoopF{L}{L}{D}\,+\frac14\;\delta_{\alpha\beta}\;\TwoLoopA{L}{L}{L}{L}\;.    
\end{equation}
For the remaining diagrams, the reduction procedure is completely analogous. One iteratively removes scalar products of momenta by using $k\cdot l=\tfrac12\,\big(k^2+l^2-(k-l)^2\big)$ and cancels propagators when possible, integrates by parts when necessary, and further uses $\int dk\,1=0$ as well as $SO(n)$ symmetry, which sets to zero all parity odd integrals.
Applying this procedure to all diagrams in \eqref{Gamma diagrams}, one is left with the two master integrals
\begin{equation}
\TwoLoopA{L}{L}{L}{L}\;,\quad \TwoLoopF{L}{L}{D}\;,
\end{equation}
while the remaining diagrams can be reduced as follows:
\begin{equation}\label{redux}
\begin{split}
\TwoLoopB{L}{L}{L}{D}&\;=\;\frac{1}{2}\;\delta_{\alpha \beta}\;\TwoLoopA{L}{L}{L}{L}\\
\TwoLoopD{L}{L}{D}{L}&\;=\;\frac{1}{2}\;\delta_{\alpha \beta}\;\TwoLoopA{L}{L}{L}{L}\;, \\
\TwoLoopE{L}{D}{L}{D}&\;=\;\frac{1}{4}\;\delta_{\alpha \beta}\;\TwoLoopA{L}{L}{L}{L}\;, \\
\TwoLoopG{D}{L}{L}&\;=\frac14\;\TwoLoopF{L}{L}{D}\,+\frac14\;\delta_{\alpha\beta}\;\TwoLoopA{L}{L}{L}{L}\;,\\
\TwoLoopH{D}{L}{D}&\;=-\frac14\;\TwoLoopA{L}{L}{L}{L}\;,\\ \TwoLoopI{D}{L}{D}{L}{D}&\;=\frac14\;\TwoLoopF{L}{L}{D}\;-\frac{1}{4}\;\delta_{\alpha \beta}\;\TwoLoopA{L}{L}{L}{L}\;,\\  \TwoLoopI{D}{D}{L}{D}{L}&\;=\frac14\;\TwoLoopF{L}{L}{D}\;+\frac{1}{4}\;\delta_{\alpha \beta}\;\TwoLoopA{L}{L}{L}{L}\;,\\ 
\TwoLoopJ{L}{D}{L}{D}{D}&\;=-\frac14\;\TwoLoopF{L}{L}{D}\;+\frac38\;\delta_{\alpha \beta}\;\TwoLoopA{L}{L}{L}{L}\;, 
\end{split}    
\end{equation}
where we have discarded the two diagrams with a blue tadpole, since they vanish in dimensional regularization. Interestingly, these are the only diagrams whose worldsheet structure is not $O(d,d)$ covariant, because of the $GL(d)$ trace $\tr\big(g^{-1}\cD^2g\big)$.

Having found the reduction, it is now time to compute the master integrals and subtract their subdivergences. The evaluation of the two master integrals trivially reduces to products of the basic tadpole $I$, which is given in \eqref{I}. While for the chain topology this can be seen already from \eqref{master1}, for the sunset we need to use $SO(n)$ symmetry to reduce the tensor integral $J_{\alpha\beta}$ to a scalar integral: $J_{\alpha\beta}=\frac1n\,\delta_{\alpha\beta}\,J^\gamma{}_\gamma$, yielding\footnote{Recall that 
$\mu^2=4\pi e^{-\gamma}\tilde\mu^2$ and that we are always discarding any finite part.} 
\begin{equation}\label{masters}
\begin{split}
\TwoLoopA{L}{L}{L}{L}\;&=I^2=\frac{1}{4\pi^2}\,\left(\frac{1}{\epsilon^2}+\frac{1}{\epsilon}\,\log\frac{m^2}{\mu^2}\right)\;, \\
\TwoLoopF{L}{L}{D}&=\frac{1}{n}\,\delta_{\alpha\beta}\,I^2 =\frac{\delta_{\alpha\beta}}{8\pi^2}\,\left(\frac{1}{\epsilon^2}+\frac{1}{\epsilon}\,\log\frac{m^2}{\mu^2}-\frac{1}{2\epsilon}\right) \;.
\end{split}    
\end{equation}
At this point a few comments are in order: while the master integral with chain topology is unique, the sunset master integral is not, depending on the positioning of the derivatives. At the massless level, the two configurations of derivatives are related by integration by parts in configuration space, as
\begin{equation}
\int dkdl\,\frac{l_{(\alpha} (k-l)_{\beta)}}{k^2l^2(k-l)^2}=-\frac12\int dkdl\,\frac{l_{\alpha} l_{\beta}}{k^2l^2(k-l)^2}\;.
\end{equation}
The infrared regularization of the two configurations, however, is not the same. While the configuration on the right-hand side (which is the one we chose) requires at least two masses as in \eqref{master4}, the configuration on the left-hand side requires at least one mass:
\begin{equation}
\int dkdl\,\frac{l_{(\alpha} (k-l)_{\beta)}}{k^2l^2(k-l)^2}\;\xrightarrow{\text{IR reg.}}\;\int dkdl\,\frac{l_{(\alpha} (k-l)_{\beta)}}{(k^2+m^2)l^2(k-l)^2} \;.  \end{equation}
While the subtracted integrals have the same poles in $\epsilon$, regardless of the IR regularization used, the way to compute them is quite different: in the configuration we have chosen, the integral can be evaluated trivially due to propagator cancellation. In the other case instead, one first computes the one-loop $l-$bubble:
\begin{equation}
\begin{split}
V_{\alpha\beta}(k)&=\int dl\,\frac{l_{(\alpha} (k-l)_{\beta)}}{l^2(k-l)^2}\\
&=\frac{1}{4\pi}\,\left(\frac{k^2}{4\pi\tilde\mu^2}\right)^{\epsilon/2}\,\frac{\Gamma^2(1+\epsilon/2)}{\Gamma(2+\epsilon)}\,\left[\frac{k_\alpha k_\beta}{k^2}\,\Gamma(1-\epsilon/2)-\frac12\,\delta_{\alpha\beta}\,\Gamma(-\epsilon/2)\right]\;,    
\end{split}
\end{equation}
which is infrared finite at non-vanishing $k$. The $k-$dependent $V_{\alpha\beta}$ is then inserted in the remaining $k-$integral. Although it is not hard to choose this method, it certainly involves nontrivial computations.

Let us now come to the $\R$ operation to remove subdivergences. This operation is recursive, allowing to subtract subdivergences at any loop order \cite{Kleinert:2001ax}. In the two-loop case, we shall proceed by following these steps:
\begin{enumerate}
\item Given a two-loop master integral, consider all possible one-loop subdiagrams obtained by cutting lines open.\\[-4mm]
\item For each one-loop subdiagram, extract the divergent part (an operation that we denote by $\cK$). This shrinks the original subdiagram to a vertex, which we denote by a cross \begin{tikzpicture} \draw[fill=white] circle (.12); \draw  node[X]{}; \end{tikzpicture}.\\[-4mm]
\item Substitute the one-loop subdiagram in the original two-loop diagram with the cross vertex found in the previous step. The substitution and insertion operation are denoted by $\star$.
\end{enumerate}
We will examine in detail the subtraction procedure for the sunset master integral, since the case of the chain master integral is simpler. First of all, the diagram \eqref{master4} has two independent subdiagrams, the first of which appearing twice:
\begin{equation}\label{sub2}
2\times\;\OneLoopA{E}{E}{L}{L}{D}{L}\qquad{\rm and}\qquad\OneLoopA{E}{E}{D}{D}{L}{L}\;.
\end{equation}
The second subdiagram is UV finite and thus does not enter the subtraction. It should be mentioned that in \eqref{sub2} the wavy blue lines still carry zero momentum, but the other external lines have to be taken with arbitrary momenta, since they belong to the two-loop vacuum diagram. In this case (and in the other master integral as well), the divergence arises only at zero external momentum and is easily computed as
\begin{equation}\label{kappasunset}
\cK\left(\OneLoopA{E}{E}{L}{L}{D}{L}\right)=\cK\int dk\,\frac{k_\alpha k_\beta}{k^2(k^2+m^2)}=\delta_{\alpha\beta}\,\cK\left(\frac{1}{n}\,I\right)=-\frac{1}{4\pi\epsilon}\,\delta_{\alpha\beta}\;,    
\end{equation}
where we notice that the masses remain in the same propagators as in the original diagram. The insertion operator $\star$ reduces in this case to multiplication by \eqref{kappasunset}. In more complicated cases, where the divergent part of a subdiagram has momentum dependence, this has to be inserted at the position of the cross vertex. In the case at hand this simply gives
\begin{equation}\label{subtraction2}
\begin{split}
2\,\cK\left(\OneLoopA{E}{E}{L}{L}{D}{L}\right)\star \CounterLoopA{L}&=-\frac{1}{2\pi\epsilon}\,\delta_{\alpha\beta}\,\CounterLoopA{L}=-\frac{1}{2\pi\epsilon}\,\delta_{\alpha\beta}\,I\\
&=\frac{1}{8\pi^2}\,\delta_{\alpha\beta}\,\left(\frac{2}{\epsilon^2}+\frac{1}{\epsilon}\,\log\frac{m^2}{\mu^2}\right)\;.
\end{split}    
\end{equation}
Subtracting \eqref{subtraction2} from the value of the sunset in \eqref{masters} finally gives the subtracted result
\begin{equation}\label{the only}
\begin{split}
\R\left[\TwoLoopF{L}{L}{D}\right]&=\TwoLoopF{L}{L}{D}-2\,\cK\left(\OneLoopA{E}{E}{L}{L}{D}{L}\right)\star \CounterLoopA{L}\\
&=-\frac{\delta_{\alpha\beta}}{8\pi^2}\,\left(\frac{1}{\epsilon^2}+\frac{1}{2\epsilon}\right)\;.
\end{split}    
\end{equation}
Applying the same procedure to the chain master integral gives
\begin{equation}\label{subtracted chain}
\begin{split}
\R\left[\TwoLoopA{L}{L}{L}{L}\right] &= \TwoLoopA{L}{L}{L}{L} -2\,\K \left( \OneLoopI{L}{L}{L}\right) \star \CounterLoopA{L}\\
&=-\frac{1}{4 \pi^2 \epsilon^2}\;.
\end{split}    
\end{equation}
Let us pause to discuss this result. First of all, a good sanity check of the computation is that the divergent terms should be local and independent of both the mass regulator $m$ and the renormalization scale $\mu$. One can see from \eqref{the only} and \eqref{subtracted chain} that this is indeed the case. More importantly, the subtracted result of the chain diagram has no simple pole $\frac{1}{\epsilon}$. Since the beta function is computed, at all loops, from the simple pole part of the counterterms (see, \emph{e.g.} \cite{Tseytlin:1986ws}), this implies that the chain diagrams do not contribute to the beta function. This is an example of a more general result: by using this direct subtraction method it can be proven \cite{Kleinert:2001ax} that no diagram with the factorized chain topology can exhibit a $\frac{1}{\epsilon}$ pole. Besides implying that the first six diagrams in \eqref{Gamma diagrams} can be discarded when computing the beta function, this also entails that any chain appearing in the reduction \eqref{redux} can be ignored as well. We will compute the contribution of the chains nonetheless, in order to display the full two-loop divergences of the effective action.

Using the subtracted values \eqref{the only} and \eqref{subtracted chain} in the decomposition \eqref{redux}, one finds the values of all the integrals appearing in \eqref{Gamma diagrams}. Pairing them with the corresponding worldsheet structures, the two-loop divergences of the effective action can be written in the form
\begin{equation}
\Gamma_{g,2l\,{\rm div}}=-\frac{1}{2\lambda}\int d^2x\,\left(\frac{1}{\epsilon^2}\,M^{\scalebox{0.6}{(2)}}_{ij}+\frac{1}{\epsilon}\,M^{\scalebox{0.6}{(1)}}_{ij}\right)\,\del^\alpha Y^i\del_\alpha Y^j \;,   
\end{equation}
where we extracted a factor of $-\frac{1}{2\lambda}$ to identify the $M^{\scalebox{0.6}{(1,2)}}_{ij}$ tensors with the counterterms. These $GL(d)$ tensors are the ${}_{ij}$ components of $O(d,d)$ matrices $\cM^{\scalebox{0.6}{(1,2)}}$ constructed from $\cS$ and $\cD$. The direct reading of the tensor structures from the diagrams in \eqref{Gamma diagrams} is in terms of $\cD^n\cS$, with $n$ up to four. This is not a good basis, since $\cD^n\cS$ has no definite parity, except for $n=0,1$.
We thus introduce a basis of independent odd structures, which we choose to be $\cD\cS$, $\cT$, $\big(\cD\cT\big)_-$ and $\big(\cD^2\cT\big)_-$, where the subscript ${}_-$ denotes the odd projection as in \eqref{projection}. All four-derivative structures, both even and odd, can be written in terms of this odd basis and $\cS$.    
The manifest parity decomposition of $\cD^n\cS$ in this basis is given, up to fourth order, by
\begin{equation}\label{parity basis}
\begin{split}
\cD\cS&=\big(\cD\cS\big)_-\;,\\
\cD^2\cS&=\big(\cD^2\cS\big)_--\cS\big(\cD\cS\big)^2\;,\quad \big(\cD^2\cS\big)_-\equiv\cT\;,\\
\cD^3\cS&=\big(\cD\cT\big)_--\big(\cD\cS\big)^3-\tfrac32\,\cS\,\big(\cD\cS\cT+\cT\cD\cS\big)\;,\\
\cD^4\cS&=\big(\cD^2\cT\big)_--2\,\cS\,\Big(\cD\cS(\cD\cT)_-+(\cD\cT)_-\cD\cS\Big)-3\,\cS\cT^2\\
&\hspace{5mm}+\cS(\cD\cS)^4-2\,\cD\cS\cT\cD\cS-\tfrac32\,\Big(\cT(\cD\cS)^2+\cT(\cD\cS)^2\Big)\;.
\end{split}    
\end{equation}
Using the decomposition \eqref{parity basis} and recalling that $\lambda=2\pi\alpha'$, the matrices $\cM^{\scalebox{0.6}{(1,2)}}$ can be finally written as
\begin{equation}\label{counterterms}
\begin{split}
\cM^{\scalebox{0.6}{(2)}}&= \frac{{\alpha'}^{\,2}}{8} \left[ \big(\cD^2\cT\big)_- - \cS \cT^2 + \frac{1}{2}\big(\cD \cS\big)^2 \cT + \frac{1}{2} \cT \big(\cD \cS\big)^2 + \frac{1}{4} \cT\, \Tr\big(\cD \cS\big)^2 \right]\;,\\
\cM^{\scalebox{0.6}{(1)}}&=\frac{{\alpha'}^{\,2}}{8}\left[ \frac{1}{2}\big(\cD \cS\big)^2 \cT + \frac{1}{2} \cT \big(\cD \cS\big)^2 + \cS \cT^2 - \frac{1}{8} \cS \big(\cD \cS\big)^2\, \Tr\big(\cD \cS\big)^2 \right]\;.
\end{split}    
\end{equation}

While the simple pole part of the counterterm determines the beta function, the higher pole term $\cM^{\scalebox{0.6}{(2)}}$ is typically used for consistency checks. Higher pole counterterms obey the so-called pole equations \cite{tHooft:1973mfk,AlvarezGaume:1981hn}, but we choose a simpler consistency check based on duality covariance, which we now illustrate. The renormalization procedure allows to relate the bare coupling $g_{0ij}$ to the renormalized one $g_{ij}$ via counterterms. If the renormalization can be made duality covariant, one should be able to relate the bare generalized metric $\cS_0$ to the renormalized one $\cS$. Demanding that both bare and renormalized generalized metrics obey the $O(d,d)$ constraint, \emph{i.e.} $\cS_0^2=1$ and $\cS^2=1$, one obtains a non-trivial constraint on $\cM^{\scalebox{0.6}{(2)}}$:
\begin{equation}\label{constraint}
\cS\cM^{\scalebox{0.6}{(2)}}+\cM^{\scalebox{0.6}{(2)}}\cS+\left(\boldsymbol{\beta}_{1l}\right)^2=0\;,    
\end{equation}
where the one-loop beta function in matrix notation is simply given by
\begin{equation}
\boldsymbol{\beta}_{1l}=-\frac{\alpha'}{2}\,\cT\;.    
\end{equation}
Had we expressed $\cM^{\scalebox{0.6}{(2)}}$ in terms of $\cD^4\cS$, $\cD^3\cS$ and $\cD^2\cS$, verifying \eqref{constraint} would be rather cumbersome. The advantage of the parity basis used to write \eqref{counterterms} is clear, in that parity even terms commute with $\cS$, while parity odd terms anti-commute, yielding immediately
\begin{equation}
\cS\cM^{\scalebox{0.6}{(2)}}+\cM^{\scalebox{0.6}{(2)}}\cS=2\,\cS\cM^{\scalebox{0.6}{(2)}}_+=-\frac{{\alpha'}^{\,2}}{4}\,\cT^2\;,    
\end{equation}
which obviously verifies \eqref{constraint}.
We conclude this section by giving the explicit form of the two-loop beta function, which is given by $\boldsymbol{\beta}_{2l}=-2\,\cM^{\scalebox{0.6}{(1)}}$ and reads
\begin{equation}\label{beta 2loop}
\boldsymbol{\beta}_{2l}=-\frac{{\alpha'}^{\,2}}{4}\left[ \frac{1}{2}\big(\cD \cS\big)^2 \cT + \frac{1}{2} \cT \big(\cD \cS\big)^2 + \cS \cT^2 - \frac{1}{8} \cS \big(\cD \cS\big)^2\, \Tr\big(\cD \cS\big)^2 \right]\;.
\end{equation}

\subsection{Field equations for target-space theory}

Having computed the two-loop beta function, the field equations \eqref{eom 2loop general} are given by
\begin{equation}\label{eom 2loop ambiguous}
\begin{split}
\cE_{p,q}&=-\frac12\,\Big(\cT-\cD\phi\cD\cS\Big)-\frac{\alpha'}{4}\,\Bigg[ \frac{1}{2}\big(\cD \cS\big)^2 \cT + \frac{1}{2} \cT \big(\cD \cS\big)^2\\ &\hspace{50mm}+\big(1+4\,q\big)\left( \cS \cT^2 - \frac{1}{8} \cS \big(\cD \cS\big)^2\, \Tr\big(\cD \cS\big)^2\right)\\
&\hspace{50mm}-4p\,\cT\,\Tr\big(\cD\cS\big)^2-4\big(p+\tfrac12\,k_2\big)\,\cD\cS\, \Tr\big(\cT\cD\cS\big) \Bigg]=0\;,    
\end{split}    
\end{equation}
where we included the ambiguity \eqref{ambiguity} and the unknown $\cW$ vector \eqref{W2}. Before comparing the equation \eqref{eom 2loop ambiguous} with the one obtained from the target-space action, let us analyze the structure of the $\alpha'$ correction. 
We recall that the two parameters $p$ and $q$ are completely arbitrary, reflecting the ambiguity in the renormalization scheme. The parameter $k_2$, on the other hand, ought to be determined once $p$ and $q$ are chosen.
Coming to the $O(d,d)$ tensor structures appearing in \eqref{eom 2loop ambiguous},
the first line  consists of parity odd terms with no traces. These terms are not affected by the ambiguity. The second line contains parity even terms, in a combination which is fully ambiguous, while the last line displays parity odd terms with a trace.

As we have discussed in the previous sections, any field equation for $\cS$ must have definite odd parity in order to be duality invariant\footnote{The  odd parity requirement is related to maintaining the constraint $\cS^2=1$ under renormalization.}, meaning that it should obey
\begin{equation}\label{parity constraint}
\cS\cE_{p,q}\cS+\cE_{p,q}=0\;.    
\end{equation}
Since the sigma model \eqref{cosmo sigma} only exhibits manifest $GL(d)$ symmetry, one does not expect \eqref{parity constraint} to hold for arbitrary renormalization schemes $(p,q)$. One can see, however, that all schemes with $q=-\frac14$ do obey \eqref{parity constraint} and are thus duality invariant. Let us stress that there are three independent parity even structures with four derivatives\footnote{From the diagrammatic expansion of the effective action one can see that only odd powers of $\cS$ can appear.}: $\cS\cT^2$, $\cS\big(\cD\cS\big)^2\Tr\big(\cD\cS\big)^2$ and $\cS\big(\cD\cS\big)^4$, but only one linear combination is ambiguous, namely $\cS\cT^2-\frac18\,\cS\big(\cD\cS\big)^2\Tr\big(\cD\cS\big)^2$. It is thus a highly nontrivial check of our computation that the only parity even terms in \eqref{beta 2loop} appear in the ambiguous combination. We shall thus choose the duality invariant scheme $q=-\frac14$ and rewrite the field equation as
\begin{equation}\label{eom 2loop parity+}
\begin{split}
\cE_{p,-\frac14}&=-\frac12\,\Big(\cT-\cD\phi\cD\cS\Big)-\alpha'\,\Bigg[ \frac{1}{8}\big(\cD \cS\big)^2 \cT + \frac{1}{8} \cT \big(\cD \cS\big)^2\\
&\hspace{50mm}-p\,\cT\,\Tr\big(\cD\cS\big)^2-\big(p+\tfrac12\,k_2\big)\,\cD\cS\, \Tr\big(\cT\cD\cS\big) \Bigg]=0\;.    
\end{split}    
\end{equation}

The equivalence of \eqref{eom 2loop parity+} with the field equation of the target-space theory is perturbative in $\alpha'$. In particular, this means that one can rewrite terms by using the lower order equations, \emph{i.e.}
\begin{equation}\label{lower eoms}
\begin{split}
\cT&=\cD\phi\cD\cS+\cO(\alpha')\;,\\
\cD^2\phi&=\big(\cD\phi\big)^2+\cO(\alpha')\;,\\
\Tr\big(\cD\cS\big)^2&=-8\,\big(\cD\phi\big)^2+\cO(\alpha')\;,
\end{split}    
\end{equation}
only committing errors of order ${\alpha'}^{\,2}$.
In particular, upon using \eqref{lower eoms} one can see that all four-derivative odd tensors reduce to two different structures:
\begin{equation}\label{on-shell values}
\begin{split}
\cT\big(\cD\cS\big)^2&=\big(\cD\cS\big)^2\cT=\cD\cS\cT\cD\cS=\cD\phi\big(\cD\cS\big)^3+\cO(\alpha')\;,\\
\cT\,\Tr\big(\cD\cS\big)^2&=\cD\cS\,\Tr\big(\cT\cD\cS\big)=-8\,\big(\cD\phi\big)^3\cD\cS+\cO(\alpha')\;,\\
\big(\cD^2\cT\big)_-&=6\,\big(\cD\phi\big)^3\cD\cS-\cD\phi\big(\cD\cS\big)^3+\cO(\alpha')\;.
\end{split}    
\end{equation}
Using the on-shell values \eqref{on-shell values}, the field equation \eqref{eom 2loop parity+} is perturbatively equivalent to the simpler form
\begin{equation}\label{eom on-shell}
\widetilde\cE_{p}=-\frac12\,\Big(\cT-\cD\phi\cD\cS\Big)-\alpha'\,\left[ \frac{1}{4}\,\cD\phi\big(\cD \cS\big)^3+4\,\big(4p+k_2\big)\,\big(\cD\phi\big)^3\cD\cS\right]=0\;.    
\end{equation}
Since the original field equation \eqref{eom 2loop parity+} does not contain the term $\big(\cD^2\cT\big)_-=\cD^4\cS+\cdots$, one can still distinguish the $O(d,d)$ structures with and without traces from the on-shell values \eqref{on-shell values}. We are now ready to compare \eqref{eom on-shell} with the field equation obtained from the target-space action
\begin{equation}\label{I0+I1}
I_0+\alpha'\,I_1=\int dt\,n\,e^{-\phi}\Big[-\big(\cD\phi\big)^2+\Tr\Big(-\tfrac18\,\big(\cD\cS\big)^2+\alpha'c_2\,\big(\cD\cS\big)^4\Big)\Big]    \;.
\end{equation}
Given that the action \eqref{I0+I1} does not contain double traces, no single trace can appear in the field equations. Using the lower order equations implies that the order $\alpha'$ correction does not contain terms $\big(\cD\phi\big)^3\cD\cS$. Demanding this to be consistent with \eqref{eom on-shell}, determines $k_2=-4p$ for a given choice of scheme $p$. Varying \eqref{I0+I1} with respect to $\cS$ yields
\begin{equation}
\begin{split}
\cE&=-\frac12\,\Big(\cT-\cD\phi\cD\cS\Big) +8\,\alpha'c_2\Big[\cT\big(\cD\cS\big)^2+\cD\cS\cT\cD\cS+\big(\cD\cS\big)^2\cT-\cD\phi\big(\cD\cS\big)^3\Big]\\
&=-\frac12\,\Big(\cT-\cD\phi\cD\cS\Big) +16\,\alpha'c_2\,\cD\phi\big(\cD\cS\big)^3+\cO({\alpha'}^{\,2})=0\;.
\end{split}    
\end{equation}
Finally, comparing the above equation with \eqref{eom on-shell} determines the $c_2$ coefficient to be 
\begin{equation}
c_2=-\frac{1}{64}\,, 
\end{equation} 
which, including the sign difference due to Euclidean signature, coincides with the known result for the bosonic string  \cite{Meissner:1996sa,Hohm:2015doa}.

\subsection{Simplifications  towards higher loops}

In the previous sections we computed the full two-loop beta function \eqref{beta 2loop}, as well as the higher pole term $\cM^{\scalebox{0.6}{(2)}}$, including ambiguities in the renormalization scheme. Here we will discuss a strategy to maximally simplify the computation, with the only goal of fixing the coefficients of the $\alpha'$ corrections.

First of all, if one is only interested in computing the beta function, all diagrams with a product topology (which at two loops consist of the chain topology) can be ignored. In our case this already reduces the number of diagrams from twelve to six. Moreover, duality invariance of the target-space equations implies that there should be a renormalization scheme (which corresponds to $q=-\frac14$ in our case) in which the beta function has definite odd parity. Assuming this to be the case allows to ignore the last two Feynman diagrams in \eqref{Gamma diagrams}, since they have the purely even structure $\cS\big(\cD\cS\big)^4$. Finally, from the analysis of \cite{Hohm:2019jgu} one knows that trace terms should be removable by field redefinitions\footnote{This is true at two and three-loop level. For higher loops, traces can be ignored nonetheless if one is only interested in determining the coefficients of the single trace terms of the target-space action.}. Assuming that traces do not contribute in determining $c_2$ allows to ignore two more diagrams with a closed blue loop in \eqref{Gamma diagrams}. At the end, $c_2$ is determined by two Feynman diagrams, which in turn depend on a single master integral, as follows:
\begin{equation}\label{Gamma relevant}
\begin{split}
\Gamma_{\rm relevant}&=-\frac{\lambda}{4}\,\R_{\frac{1}{\epsilon}}\left(\TwoLoopF{L}{L}{D}\right)_{\alpha\beta}\int d^2x\,\Big(\cD^2\cS\cS\cD^2\cS\Big)_{-\,ij}\,\del^\alpha Y^i\del^\beta Y^j\\
&\hspace{5mm}-\lambda\,\R_{\frac{1}{\epsilon}}\left(\TwoLoopG{D}{L}{L}\right)_{\alpha\beta}\int d^2x\,\Big(\big(\cD\cS\big)^2\cD^2\cS\Big)_{-\,ij}\,\del^\alpha Y^i\del^\beta Y^j \\[2mm]
&=-\frac{\lambda}{4}\,\R_{\frac{1}{\epsilon}}\left(\TwoLoopF{L}{L}{D}\right)_{\alpha\beta}\int d^2x\,\Big(\cD^2\cS\cS\cD^2\cS+\big(\cD\cS\big)^2\cD^2\cS\Big)_{-\,ij}\,\del^\alpha Y^i\del^\beta Y^j\\[2mm]
&=\frac{\lambda}{64\pi^2\epsilon}\int d^2x\,\Big(\cD^2\cS\cS\cD^2\cS+\big(\cD\cS\big)^2\cD^2\cS\Big)_{-\,ij}\,\del^\alpha Y^i\del_\alpha Y^j\;,
\end{split}    
\end{equation}
where by the subscript ${_-}$ we denoted the odd projection and by $\cR_{\frac{1}{\epsilon}}$ we meant to discard higher poles. We have also used the reduction \eqref{redux} in terms of master integrals, and the subtracted value \eqref{the only}. The odd projection of the above $O(d,d)$ tensor is given by
\begin{equation}
\Big(\cD^2\cS\cS\cD^2\cS+\tfrac12\,\cD^2\cS\big(\cD\cS\big)^2+\tfrac12\,\big(\cD\cS\big)^2\cD^2\cS\Big)_-=-\tfrac12\,\Big(\cT\big(\cD\cS\big)^2+\big(\cD\cS\big)^2\cT\Big)   \;,
\end{equation}
Finally, \eqref{Gamma relevant} yields the beta function
\begin{equation}
\boldsymbol{\beta}_{2l}=-\frac{{\alpha'}^{\,2}}{8}\,\Big(\cT\big(\cD\cS\big)^2+\big(\cD\cS\big)^2\cT\Big)=-\frac{{\alpha'}^{\,2}}{4}\,\cD\phi\big(\cD\cS\big)^3+\cO\big({\alpha'}^{\,3}\big)\;,
\end{equation}
which is indeed the relevant part of \eqref{beta 2loop}.

\section{Summary and Outlook}

This paper is a continuation of the program to determine  the higher-derivative 
$\alpha'$ corrections of classical string theory in a duality or $O(d,d;\mathbb{R})$ invariant way. 
The $O(d,d;\mathbb{R})$ symmetry emerges for any closed string theory when dimensionally reduced along 
$d$ dimensions (or, equivalently, when considering backgrounds with $d$-dimensional translation invariance). 
Accordingly, the traditional method to compute the $O(d,d;\mathbb{R})$ invariant $\alpha'$ corrections would be 
to start from the known corrections in ten or 26 dimensions and to dimensionally reduce. 
The challenge here is that displaying  $O(d,d;\mathbb{R})$ in its conventional form requires 
laborious field redefinitions  \cite{Meissner:1996sa,Hohm:2015doa}. It is thus greatly desirable to find a procedure that directly 
determines the $O(d,d;\mathbb{R})$ invariant corrections in the dimensionally reduced theory, say 
in the cosmological context discussed here. 

We revisited the Tseytlin formulation and its generalizations, which have the advantage of being 
already manifestly $O(d,d;\mathbb{R})$ invariant. Unfortunately, the presence of chiral bosons 
and the corresponding lack of manifest Lorentz invariance of the action complicates the computation of 
the beta functions significantly, especially beyond one-loop order, as explained in the main text. 
In this paper we introduced an alternative method, which is based on the conventional Polyakov action 
of the worldsheet theory, applied to cosmological backgrounds. 
 Despite not being  $O(d,d;\mathbb{R})$ invariant we found a surprisingly efficient procedure to determine 
 the duality invariant beta functions and tested it by computing the first two non-trivial physical coefficients 
 in the cosmological $\alpha'$ expansion, finding perfect agreement with the literature   \cite{Meissner:1996sa,Hohm:2015doa,Hohm:2019jgu}. 
 
 This new method can be extended to higher loops, and the details of the three-loop case will be presented 
 in a separate paper. Of course, the complications grow sharply with the number of loops, but 
 it is quite conceivable that with a suitable automation of the computation one could eventually push 
 it beyond the $\alpha'^3$  corrections that so far are the state-of-the-art \cite{Codina:2020kvj,Codina:2021cxh}. 
 The real goal, however, of determining the 
 $\alpha'$-complete equations, whose general form is known thanks to the classification in \cite{Hohm:2019jgu}, 
 is still out of reach. One may hope that novel geometric ideas, as present in double field theory, may 
 eventually help in overcoming these limitations.

\subsection*{Acknowledgements} 

We would like to thank Felipe Diaz-Jaramillo and Arkady Tseytlin for useful discussions and correspondence.

This work is funded   by the European Research Council (ERC) under the European Union's Horizon 2020 research and innovation programme (grant agreement No 771862)
and by the Deutsche Forschungsgemeinschaft (DFG, German Research Foundation), ``Rethinking Quantum Field Theory", Projektnummer 417533893/GRK2575.

%\newpage
\bibliography{BrokenP.bib}
\bibliographystyle{utphys}

\end{document}